\documentclass[manuscript,nonacm]{acmart}
\usepackage{graphicx}
\usepackage{subcaption}
\usepackage{multirow}
\usepackage{framed}
\usepackage{xcolor}

\usepackage{hyperref}

\definecolor{shadecolor}{gray}{0.95}

\newcounter{acmbox}
\newenvironment{acmshadedbox}[2]{%
  \refstepcounter{acmbox}\label{#2}%
  \begin{shaded}%
  \noindent\textbf{#1}\par\medskip
}{%
  \end{shaded}%
}

\newcommand{\cell}[1]{\begin{minipage}[t]{\linewidth}\raggedright #1\end{minipage}}

\newif\ifshowchanges
\showchangestrue

\ifshowchanges
  
\else
  
\fi

\AtBeginDocument{%
  
    }

\copyrightyear{2026}
\acmYear{2026}
\setcopyright{cc}
\setcctype{by}
\acmConference[CHI '26]{Proceedings of the 2026 CHI Conference on Human Factors in Computing Systems}{April 13--17, 2026}{Barcelona, Spain}
\acmBooktitle{Proceedings of the 2026 CHI Conference on Human Factors in Computing Systems (CHI '26), April 13--17, 2026, Barcelona, Spain}
\acmDOI{10.1145/3772318.3790608}
\acmISBN{979-8-4007-2278-3/2026/04}

\begin{document}

\title[Creating and Evaluating Personas Using Generative AI]{Creating and Evaluating Personas Using Generative AI: A Scoping Review of 81 Articles}

\author{Danial Amin}
\email{danialam@uwasa.fi}
\orcid{0009-0000-7597-2267}
\affiliation{%
  \institution{University of Vaasa}
  \city{Vaasa}
  \country{Finland}
}
\author{Joni Salminen}
\email{jonisalm@uwasa.fi}
\affiliation{%
  \institution{University of Vaasa}
  \city{Vaasa}
  \country{Finland}
}




\author{Farhan Ahmed}
\email{farhan.ahmed@student.uwasa.fi}
\affiliation{%
  \institution{University of Vaasa}
  \city{Vaasa}
  \country{Finland}
}

\author{Sonja M.H. Tervola}
\email{sonja.tervola@aalto.fi}
\affiliation{%
  \institution{Aalto University}
  \city{Espoo}
  \country{Finland}
}

\author{Sankalp Sethi}
\email{sankalpsethi@arizona.edu}
\orcid{0009-0006-8708-0876}
\affiliation{%
  \institution{University of Arizona}
  \city{Tucson}
  \country{USA}
}

\author{Bernard J. Jansen}
\email{jjansen@acm.org}
\affiliation{%
  \institution{Qatar Computing Research Institute, Hamad Bin Khalifa University}
  \city{Doha}
  \country{Qatar}
}

\renewcommand{\shortauthors}{Amin et al.}


\begin{abstract}
As generative AI (GenAI) is increasingly applied in persona development to represent real users, understanding the implications and limitations of this technology is essential for establishing robust practices. This scoping review analyzes how 81 articles (2022-2025) use GenAI techniques for the creation, evaluation, and application of personas. 
The articles exhibited good level of reproducibility, with 61\% of articles sharing resources (personas, code, or datasets). Furthermore, conversational persona interfaces are increasingly provided alongside traditional profiles. However, nearly half (45\%) of the articles lack evaluation, and the majority (86\%) use only GPT models. In some articles, GenAI use creates a risk of circularity, in which the same GenAI model both generates and evaluates outputs. Our findings also suggest that GenAI seems to reduce the role of human developers in the persona-creation process. To mitigate the associated risks, we propose actionable guidelines for the responsible integration of GenAI into persona development.
\end{abstract}

\ccsdesc[500]{Human-centered computing~HCI theory, concepts and models}
\ccsdesc[300]{Human-centered computing~User models}

\keywords{Generative AI, LLM, Personas, Persona Development}


\maketitle

\section{Introduction}
User personas (`personas' for short) are fictitious characters representing archetypal real users of a system, product, or service presented in a humanized form \cite{cooper1999inmates}. Since their introduction into human-computer interaction (HCI), personas have been adopted to support user-centered design (UCD) in domains such as software development \cite{blomquist2002personas,aoyama2005persona,adlin2001fake}, design \cite{lee2010designing,duda2018personas,chen2011modeling}, and healthcare \cite{hendriks2013designing,hogberg2008representing,gonzalez2018personas}. Personas help system designers focus on user needs throughout the design process by providing empathetic and memorable representations of core user segments \cite{pruitt2010persona}. As such, personas are a critical component of HCI research, user experience (UX) studies, and UCD of systems and technology.

\begin{figure}[ht]
  \centering
  \begin{subfigure}[b]{0.48\linewidth}
    \centering\includegraphics[width=\linewidth]{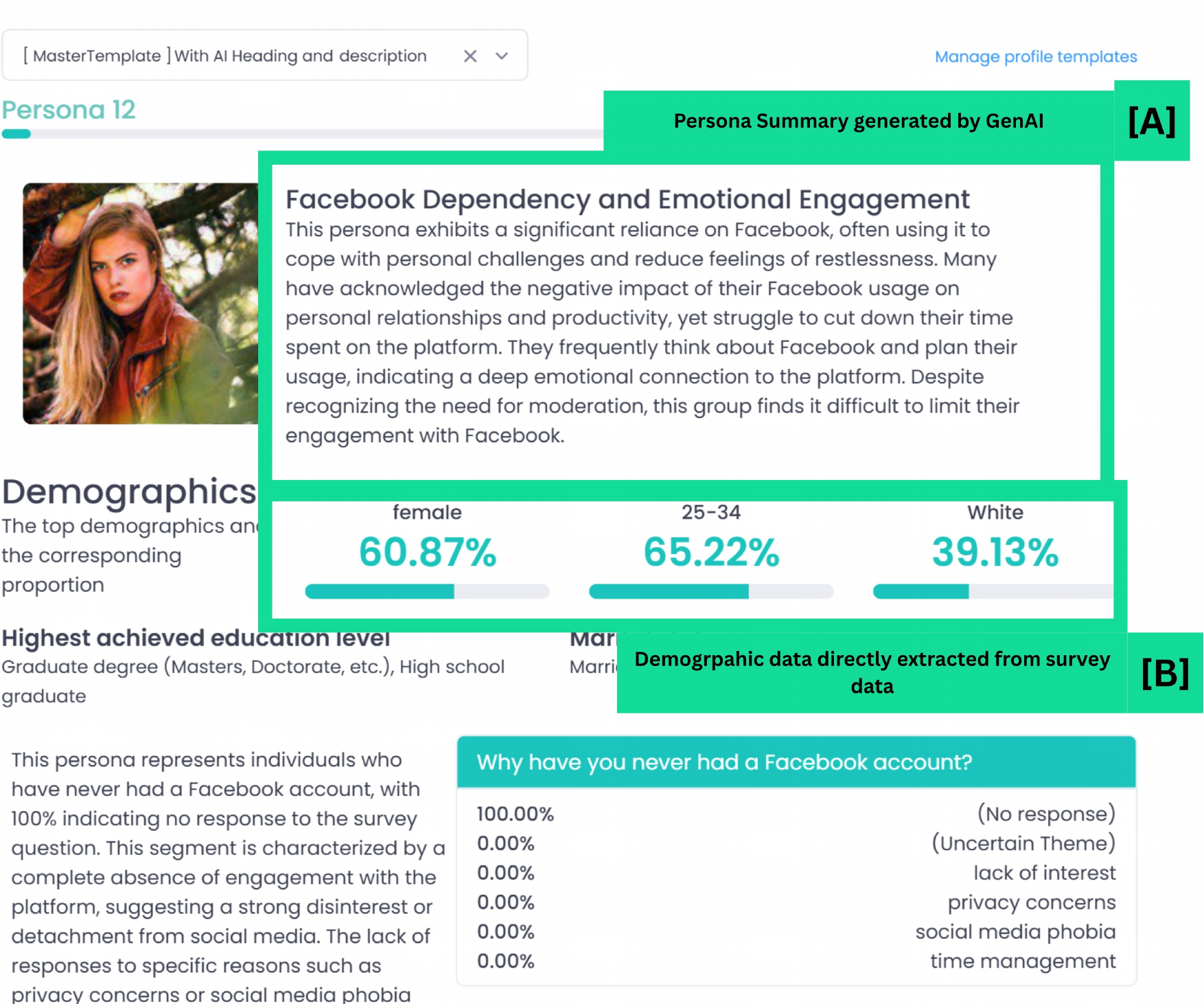} 
    \subcaption{}
    \label{fig:persona-s2p}
  \end{subfigure}
  \begin{subfigure}[b]{0.48\linewidth}
    \centering
    \includegraphics[width=\linewidth]{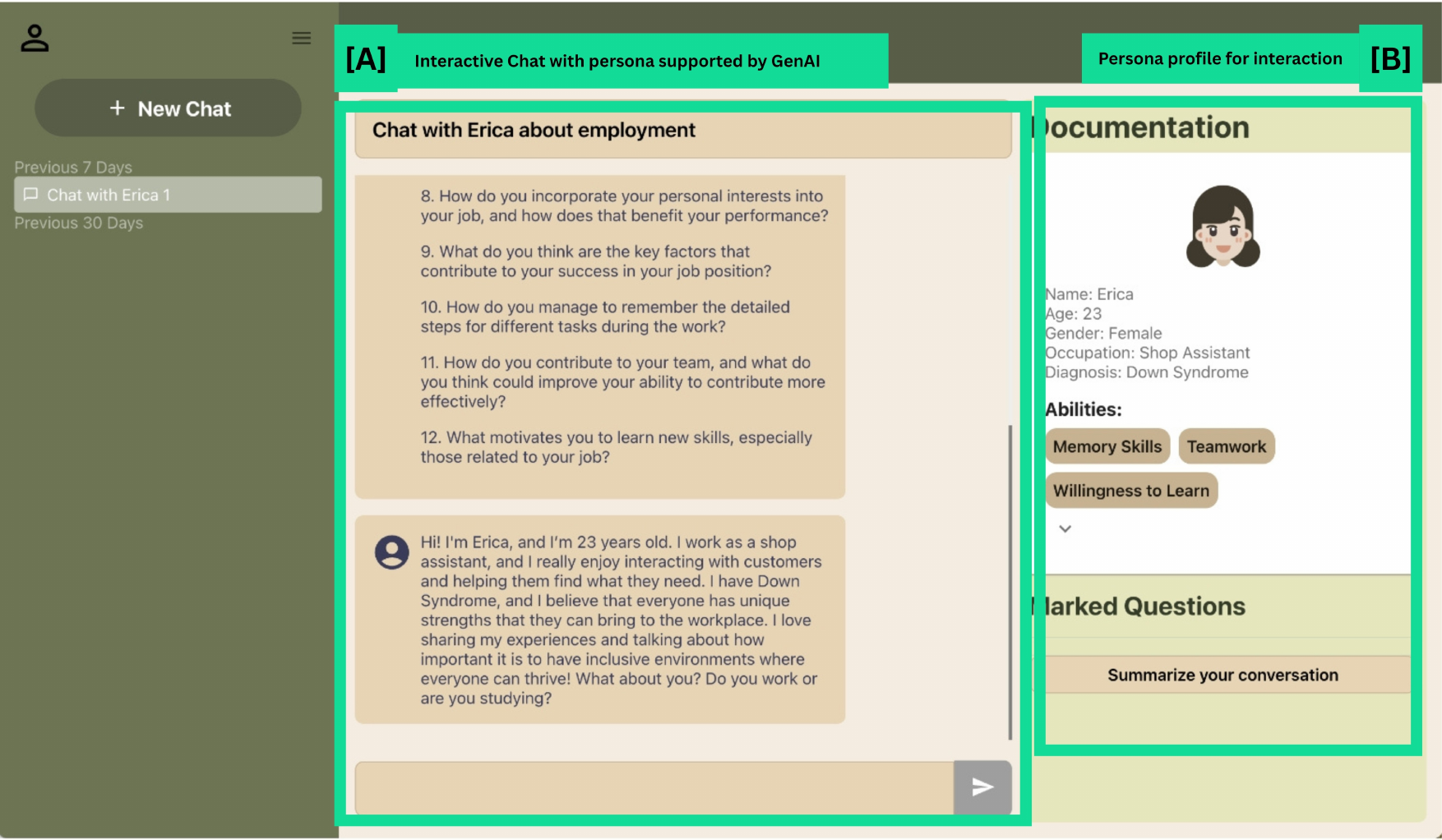} 
    \caption{}
    \label{fig:persona-l}
  \end{subfigure}
  \caption{Examples of GenAI personas. The persona in \textbf{(a)} is generated by Survey2Persona \cite{jung2025personacraft}, \textbf{(b)} is generated by Persona-L \cite{sun_persona-l_2025}. These systems illustrate that LLMs are increasingly used for generating persona profiles, but also to afford decision-makers access to dialogue with the personas. The figures are annotated to explain different components of GenAI personas. The reader is encouraged to zoom in for better readability.}
  \label{fig:personas}
  \Description{Two AI-generated persona examples. Panel A shows a persona profile for a young woman named after Survey2Persona system, displaying demographic information, personality traits, and behavioral characteristics in a structured format with percentages and data visualizations. Panel B shows a Persona-L generated interface featuring a professional woman's profile with chat functionality, demonstrating how LLMs enable both persona profiles and dialogue access for decision-makers.}
\end{figure}

Persona development methods were originally based on qualitative data collection with manual analysis, which can be time-consuming, resource-intensive, and prone to staling \cite{salminen2021survey}. However, generative artificial intelligence (GenAI), particularly large language models (LLMs), creates opportunities for automating persona development \cite{huang_exploring_2024,shin_understanding_2024,jung2025personacraft}. 
Automatic persona development is the ability to generate personas without (or minimal) human involvement and can overcome the challenges of manual persona development. This is why GenAI is being researched to support persona development. These GenAI technologies can be applied to various stages of persona development, and the personas created by such a process are called GenAI personas, which will be the focus of this study (see Figure \ref{fig:personas} for examples).

Although various articles have applied GenAI in persona development \cite{salminen_deus_2024,shin_understanding_2024,schuller_generating_2024,jung2025personacraft}, there is limited systematic synthesis of how GenAI technologies are applied in persona development practices. This lack of knowledge raises serious questions for the HCI community regarding the use of GenAI for persona creation.

Moreover, the application of GenAI in persona development suffers from a lack of standardization and best practices, 
there is yet no clear consensus on optimal approaches. The application of GenAI in persona development is currently uncharted territory, making it difficult for persona developers to make informed decisions about adopting GenAI technologies. The evolution of GenAI technologies has also increased the challenges faced by persona developers, as discussed in detail by Amin et al. \cite{amin2025generative}.
Similar to the development of data-driven personas (see a review \cite{salminen2021survey}), we can observe varying methodological choices and technical implementations. These choices involve a variety of factors, such as the selection of LLM family (e.g., GPT or Claude or Llama), versions of LLMs (e.g., GPT 3.5 or GPT 4.0), hyperparameters (i.e., parameters that impact the behavior of the model \cite{Jansen2021TheEffect}), and prompts (i.e., instructions given to the LLM to perform certain tasks). Although researchers have suggested guidelines for persona development in general \cite{hayhanen_why_2025,Jansen2021Strengths,cooper1999inmates,nielsen_template_2015,pruitt2003personas}, \textbf{how researchers specifically use GenAI in persona development remains unclear.} 

Another domain that is prone to challenges is the evaluation of GenAI personas. Traditional persona validation methods generally focused on user perceptions of personas \cite{Jansen2021,salminen_persona_2020}, but GenAI personas require additional consideration of factors, such as output consistency, prevention of hallucinations (the ability of LLMs to fabricate information), and prompt reliability \cite{sattele2024generating,lazik_impostor_2025}. Moreover, integrating GenAI raises questions about evaluating the quality of GenAI-crafted versus human-crafted personas \cite{amin2025generative}. Evaluation approaches vary widely, from automated metrics to user studies, \textbf{without a clear understanding of how researchers validate and evaluate GenAI personas.} 

The ethical aspects of using GenAI technologies may extend beyond traditional concerns about persona development. Though prior work has addressed issues of stereotyping and representation in personas \cite{turner2011stereotyping, goodman-deane_evaluating_2018,goodman-deane_developing_2021}, it appears that GenAI amplifies certain challenges, such as algorithmic bias, data privacy, and transparency \cite{amin2025generative,gupta_bias_2023,prpa_challenges_2024,hamalainen_evaluating_2023}. These challenges may become particularly acute as personas represent diverse user groups and influence design decisions that affect these populations, prompting \textbf{the need to understand how researchers factor in ethical concerns when using GenAI in persona development.} 

Against the backdrop of these research challenges, we put forth three research questions (RQs):

\begin{itemize}
    \item \textbf{RQ1:} \textit{How are GenAI technologies used in persona development?}
    \item \textbf{RQ2:} \textit{How are GenAI personas evaluated?}
    \item \textbf{RQ3:} \textit{What ethical considerations are associated with GenAI personas?}
\end{itemize}

Building on persona research in HCI \cite{pruitt2010persona,salminen2021survey} and following guidelines for literature reviews in the HCI field \cite{kitchenham2004procedures}, this scoping review analyzes 81 articles published between 2022 and 2025 on the application of GenAI for persona development. A scoping review generally maps the current scope, key concepts, and knowledge gaps \cite{arksey2005scoping}, and is suitable for studying GenAI personas given this domain's rapidly evolving nature. To this end, we identify trends, gaps, and opportunities, and we provide recommendations to practitioners concerning the integration of GenAI into persona development workflows. 
Considering the proliferation of GenAI tools and their application in user research and design processes, this work addresses a timely and relevant topic for the HCI community.

\section{Previous Reviews and Research Gap}

In UX/HCI practice, personas serve as shared reference points that help designers empathize with users and make decisions aligned with user needs \cite{pruitt2010persona}. Persona development traditionally involves collecting user data through research methods (interviews, surveys, analytics), identifying distinct user segments through analysis, and crafting humanized narrative profiles that represent these segments \cite{cooper1999inmates,salminen2021survey}. At first glance, GenAI appears to strengthen \textit{automatic persona generation} by exhibiting capability in most of these above-mentioned tasks. Automatic persona development has existed in the HCI domain for more than a decade \cite{jung2018automatic,Jung2017PersonaGeneration,Mijac2018ThePotential}. However, the limitations of automated personas have been apparent, especially when analyzing text records of users and writing up persona narratives \cite{salminen2023nlp}, which is a key activity in many persona creation approaches \cite{nielsen_personas_2019}. Algorithms and models prior to LLMs struggled with interpretive tasks, such as writing persona narratives or labeling user segments \cite{Jansen2021Strengths}, which has limited the potential for automatic persona generation, as extensive manual interventions were required in conjunction with data science methodologies \cite{Jansen2020FromFlat,Mijac2018ThePotential}. 

Due to the limited interpretative capabilities of previous AI models, researchers have applied rule-based systems, such as dynamic templates with predefined fields for dynamically inserted information \cite{nielsen_template_2015}. In contrast, LLMs' interpretative capabilities offer possible solutions to the major challenges of contextually interpreting and summarizing textual data about people \cite{jung2025personacraft}. However, questions remain and the best practices of LLMs in persona development remain unknown, including the ideal techniques and processes to follow \cite{salminen_deus_2024}. 

There are also practical hurdles; for example, there is no clear concept of human-AI collaboration in persona development, even though researchers seem to agree on the need to maintain ``human in the loop'' instead of forfeiting all decision-making to GenAI \cite{shin_understanding_2024}. Interestingly, human in the loop is a feat that is also considered valuable in persona theory, such that involving stakeholders that are intended to use the personas in the persona creation process supports the development of favorable attitudes toward the persona technique \cite{neate_co-created_2019,fuglerud_co-creating_2020}. 
Additionally, it is not evident how well HCI researchers are aware of the risks and challenges associated with implementing this GenAI for developing personas, including ethical concerns such as reinforcing existing systemic biases or marginalizing underrepresented groups when lacking real data on cultural contexts \cite{amin2025generative}. 

These lingering concerns call for a literature review that investigates existing practices, methodologies, and emerging patterns in GenAI persona development.
To this end, reviews of persona research have addressed quantitative personas \cite{salminen_literature_2020}, data-driven personas \cite{salminen2021survey}, personas for social impact \cite{guan_leveraging_2023}, and persona design applications \cite{salminen_use_2022}, among other areas. Although these reviews provide valuable insights into persona development methods, they do not specifically address the integration of GenAI technologies into persona development. We will discuss in Section \ref{discussion} how our findings are positioned in the continuum of persona research, specifically in conjunction with previous reviews. Given the growing adoption of GenAI in persona development \cite{shin_understanding_2024,hamalainen_evaluating_2023,salminen_deus_2024,choi_proxona_2025,prpa_challenges_2024,sun_persona-l_2025}, a scoping review of this emerging approach is needed and thus presented. 

\section{Methodology}

Following previous work in HCI \citep{bowman2023using,webber2023engaging,nunes2022scoping} that deals with the ramifications of novel technologies, we opted for a scoping review to address our RQs. Scoping reviews map the breadth of research in emerging areas, identify key concepts, and examine how research is conducted \cite{arksey2005scoping,levac2010scoping}.
We conducted a literature search in five academic databases: \textit{(1) ACM Digital Library (ACMDL), (2) IEEE Xplore, (3) Web of Science (WoS), (4) Scopus, and (5) arXiv}. The search focused on identifying articles at the intersection of GenAI and persona development. Our methodology consisted of (1) query development and search execution, (2) screening of articles based on inclusion/exclusion criteria, (3) backward and forward snowball sampling, (4) quality assessment, and (5) data extraction and synthesis.
The search process began with developing keyword groups related to GenAI and persona development, involving iterations of query refinement and implementing database-specific adaptations. Through screening, we identified 90 relevant articles of the 885 total found in the databases. Additional snowball sampling yielded 27 potentially relevant articles. 
Three additional articles from snowballing met our inclusion criteria, resulting in a final corpus of 93 articles, of which 12 (12.3\%) were excluded in the detailed screening, leaving a total of 81 (9.15\%) articles included in the corpus. The details of the search strategy, the screening process, and the analysis are described in the following subsections.

\subsection{Search Query Creation}
The creation of the search query focused on identifying relevant articles at the intersection of GenAI and persona development, with terms from both domains (see Table A1 in the online appendix). These terms were selected from prominent articles in the field, personal domain expertise, and LLM suggestions (Claude, ChatGPT, Llama). All searches were performed in the \texttt{title}, \texttt{abstract}, and \texttt{author keywords} fields in all databases. We searched all articles until August 25, 2025.
Initial searches revealed a high false positive rate when using wildcard operators (e.g., ``persona*''), with unrelated terms such as ``personality analysis''. We used phrase matching with exact terms instead (i.e., including ``persona'' and ``personas'' in the query explicitly). 

The search strings required database-specific adaptations to accommodate varying syntax requirements and technical limitations. For ACMDL, we used the independent field tags for different items (i.e., \texttt{[Abstract:]} field tag for abstract, \texttt{[Title:]} field tag for title, and \texttt{[Keyword:]} field tag for keywords). For IEEE Xplore, we searched the title, abstract, and keywords fields using the \texttt{All Metadata} field. For Scopus, we covered title, abstract, and keywords using the \texttt{TITLE-ABS-KEY} field specification. For WoS, we used the \texttt{Topic (TS)} field, which includes title, abstract, and keywords. Scopus required proximity operators (\texttt{W/5}), while Web of Science needed simplified syntax for compatibility.

The exact search string used across databases with database-specific syntax adaptations is provided in Section \ref{query:llm-personas}. Boolean operators combined the keyword groups using ``AND'' relationships between different concept groups, such as GenAI terms AND persona terms. We allowed ``OR'' within each group to capture terminological variations.

The complete evolution of search strings across five iterations, including abandoned variants, is documented in Section A1 of the online appendix\footnote{\href{https://osf.io/au5g6/?view_only=29c0a9a1251b43a186e1359ac9cd6a10}{https://osf.io/au5g6/?view\_only=29c0a9a1251b43a186e1359ac9cd6a10}}. 
Each string revision was tested by verifying the capture of relevant articles while excluding clearly irrelevant ones. The final query structure balanced breadth and specificity in identifying articles.

\subsection{Selection of Databases}
We investigated five academic databases for our scoping review based on their relevance and coverage. We created different versions for each of the databases (see Section A2 of the online appendix). ACMDL was chosen as the primary venue for HCI research, providing comprehensive coverage of persona development work in computing contexts. We excluded ACM Guide of Computing Literature to avoid duplication with ACMDL. IEEE Xplore was selected for its strong focus on technical implementations, covering engineering aspects of GenAI applications in persona development. WoS and Scopus were included to ensure broad interdisciplinary coverage, as persona research spans multiple HCI-related fields, including design, marketing, and social sciences.

Finally, \textit{arXiv} was included to capture recent developments in GenAI applications to personas, particularly given the rapid progress of LLMs. The inclusion of preprints from arXiv follows established practices in rapidly evolving GenAI research domains to reduce the risk of excluding articles with negative or null findings, particularly in fields where peer review lag could exclude cutting-edge developments~\cite{naveed2025comprehensive,sengar2025generative}. We only included preprints from the past two years to focus on recent developments. 

During deduplication, we prioritized entries based on the quality of the database metadata and the relevance of the domain. ACMDL entries were retained over other sources, followed by IEEE Xplore, WoS, and finally Scopus. 

\begin{acmshadedbox}{Search Query}{query:llm-personas}
(``generative artificial intelligence'' OR ``large language model'' OR ``large language models'' OR LLM OR LLMs OR ``generative model'' OR ``generative models'' OR GPT OR ``generative pre-trained transformer'' OR ``foundation model'' OR ``foundation models'' OR ``ChatGPT'' OR ``Gemini'' OR ``LLAMA'' OR ``Claude'') AND (``data-driven persona'' OR ``data-driven personas'' OR ``quantitative persona'' OR ``quantitative personas'' OR ``user persona'' OR ``user personas'' OR ``LLM-generated persona'' OR ``LLM-generated personas'' OR ``buyer persona'' OR ``buyer personas'' OR ``persona'' OR ``personas'' OR ``marketing persona'' OR ``marketing personas'' OR ``design persona'' OR ``design personas'' OR ``persona development'' OR ``persona creation'' OR ``persona generation'')
\end{acmshadedbox}

\subsection{Article Collection and Screening}
\begin{figure}
    \centering
    \includegraphics[width=\linewidth]{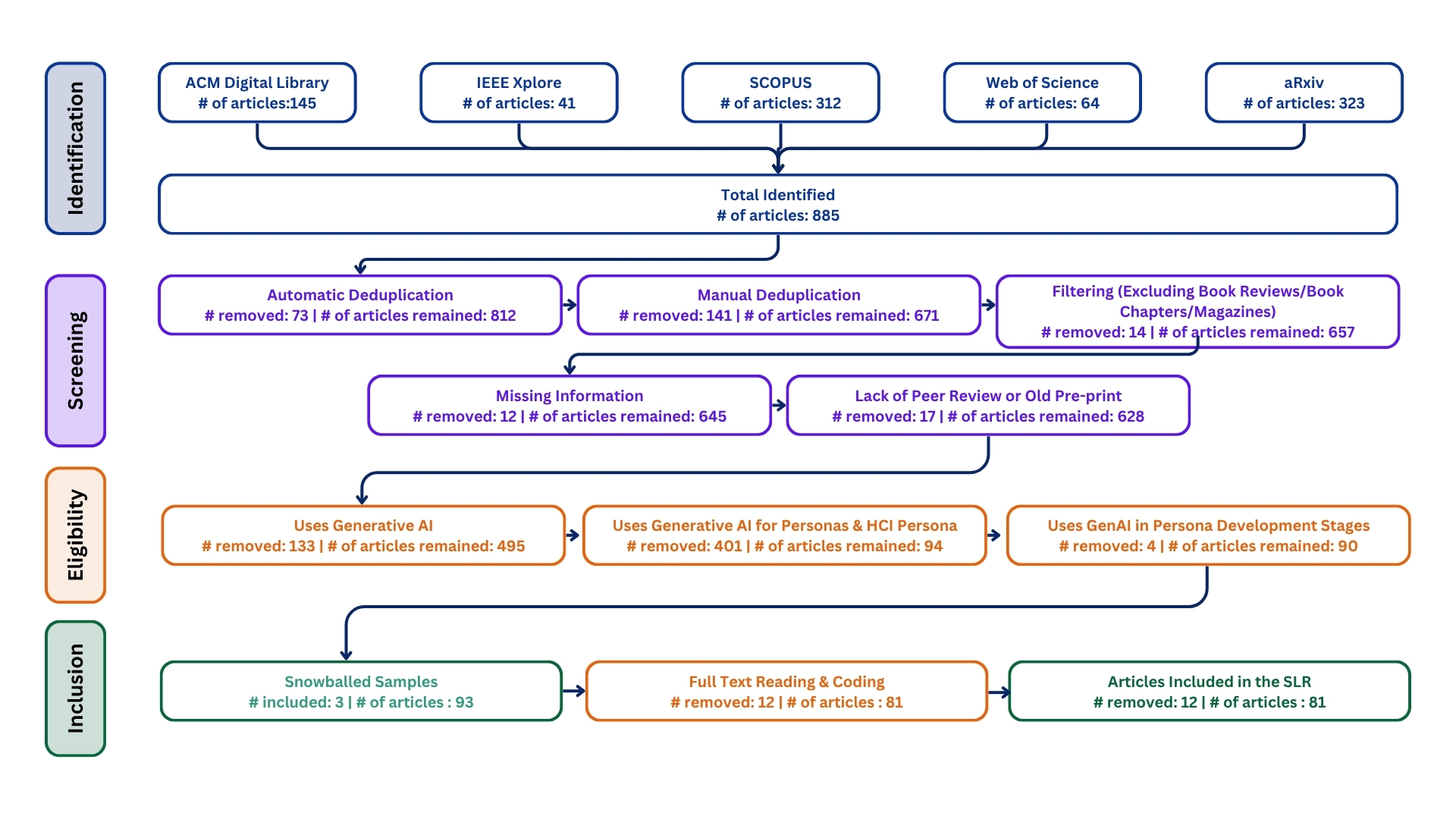}
    \caption{PRISMA diagram describing the literature collection process. We followed the standard protocols in scoping literature reviews \cite{nunes2022scoping,bowman2023using,webber2023engaging} to collect our literature base.}
    \Description{PRISMA flow diagram showing the systematic literature review process. The diagram tracks manuscript collection from five databases (ACM Digital Library, IEEE Xplore, SCOPUS, Web of Science, and arXiv) through identification, screening, eligibility assessment, and inclusion stages. Starting with 885 total manuscripts, the process resulted in 81 final manuscripts after deduplication, screening for relevance, and quality assessment.}
    \label{fig:prisma}
\end{figure}

The article collection and selection process followed the PRISMA guidelines \cite{page2021prisma}, as illustrated in Figure~\ref{fig:prisma}. The initial database search yielded 885 articles: 64 from WoS, 312 from Scopus, 145 from ACMDL, 41 from IEEE Xplore, and 323 from arXiv.
The screening process consisted of multiple stages. First, we performed automatic deduplication using reference management software, removing 73 articles, bringing the total to 812. This was followed by manual deduplication that identified 141 additional duplicates due to minor variations in article titles (e.g., case changes, encoded characters, missing information across databases). After removing conference reviews (n = 8), books and book chapters (n = 6), and other nonrelevant document types (n = 12), 645 articles remained.
The next phase involved multiple screening assessments of the 645 articles. This included verification of English language requirements (0 removed) and peer review status (17 removed). The subsequent screening for the GenAI application removed 133 articles, and the alignment assessment of the definition of HCI and the use of GenAI in personas eliminated 401 articles. Additionally, four articles were removed during the assessment stage, leaving 90 articles.

The snowball sampling process identified 27 potentially relevant articles. The snowball sample was identified by the Google Scholar Alerts service during the research period. Of these, 8 were already in our initial sample, and 5 were already in the final corpus. After screening for English language (1 removed), peer review status (0 removed), GenAI use (3 removed), and alignment of HCI definition (6 removed), 4 articles remained. One article was removed during the assessment stage, resulting in 3 additional articles for our corpus.

\subsubsection{Inclusion Criteria}
Articles were included if they met all the following criteria:

\begin{enumerate}
\item[\textbf{I1.}] Written in English to ensure accurate analysis
\item[\textbf{I2.}] Either peer-reviewed publications or preprints from the past two years
\item[\textbf{I3.}] Demonstrated application of generative AI in persona development
\item[\textbf{I4.}] Aligned with the HCI definition of personas as user representations for design. HCI definition of personas focuses on the representation of a group of users.
\item[\textbf{I5.}] Presented empirical research through qualitative, quantitative, or mixed methods
\item[\textbf{I6.}] Addressed at least one stage of persona development: data collection, analysis, generation, evaluation, or practical application
\end{enumerate}

\subsubsection{Exclusion Criteria}
Articles were excluded if they met any of the following criteria:

\begin{enumerate}
\item[\textbf{E1.}] Non-academic content: letters, editorials, book reviews, meeting abstracts, news items
\item[\textbf{E2.}] Articles only briefly mention personas or GenAI without substantial implementation
\item[\textbf{E3.}] Articles using GenAI for non-persona-related work
\item[\textbf{E4.}] Research using the term ``persona'' in non-user-representation contexts
\item[\textbf{E5.}] Articles using personas solely to enhance conversational agents' personality or behavior style.
\item[\textbf{E6.}] Purely theoretical discussions without implementation
\item[\textbf{E7.}] Opinion pieces lacking research methodology
\item[\textbf{E8.}] Duplicate publications of the same study
\end{enumerate}

The criterion \textbf{I4} is considered because research uses `persona' in at least two different meanings. HCI describes personas as a representation of a group of users \cite{pruitt2010persona,cooper1999inmates,salminen2021survey}, whereas natural language processing (NLP) refers to persona as the personality, role, attributes, or collection of design choices that are made for the conversational agent \cite{hwang_applying_2021,pradhan_2021_google,kimm_2019_designing}. For instance, among the articles excluded under this criterion, articles \cite{tao2024rolecraft,li_steerability_2024,kosenko2024krgp,kamruzzaman2024woman} discuss personas for LLMs and conversational agents as personality configurations (e.g., ``helpful assistant'' or ``creative writer'') rather than user group representations for design purposes. Similarly, the criterion \textbf{E5} differs from the conversational persona format (21.0\% of included articles, see Section \ref{narrative-output}). The included articles involving conversational personas, use conversational interfaces to represent user groups for design decisions, whereas the excluded articles use personas to configure chatbot's personalities without the intention of representing real users or user groups.

Through this screening process, we identified a corpus of 93 articles (90 from the database search + 3 from snowball sampling) for detailed analysis. Of these, 12 were removed from the full text reading stage (see details in Section \ref{sec:data_extraction_process}), so the final number of articles reviewed was 81 (9.15\%).

\subsection{Data Extraction}

\subsubsection{Data Extraction Book}
To systematically analyze the selected articles, we developed a data extraction coding book with 35 variables in six categories (see Table A3 in the online appendix): (1) basic information, (2) resources, (3) research methodology, (4) persona development, (5) GenAI / LLM usage, and (6) results and discussion. The coding book was developed by the Principal Investigator (PI) and Co-Principal Investigator (Co-PI). We reviewed the relevant coding books created for similar literature reviews on personas and included the necessary variables for our research context. We pilot tested the coding book by coding five articles at random and revised the coding book to address any challenges or ambiguities in the instructions. We also mapped the extracted variables to the RQs and identified under/over utilized variables (see Table A2 in the online appendix). We defined specific extraction instructions and value formats for each variable to ensure consistent coding across articles (see Table A3 in the online appendix). The codebook included six categories, including the basic information category, which captures publication details, and the resources category, which tracks the availability of artifacts such as code and datasets. Research methodology concerns approaches and empirical methods; persona development focuses on creation processes; and GenAI/LLM usage details how GenAI technologies are employed. Finally, the results and discussion category captures the findings, limitations, and future directions. Coders applied the codebook to each article in the corpus. For the 25 categorical variables, coders selected from predefined options listed in the codebook (see Table A3 in the online appendix). For the 10 open-ended variables, four coders independently coded responses (with 10\% overlap for agreement calculation). The open-ended variables were then coded thematically into different themes, which were jointly developed by the PI and Co-PI.

\subsubsection{Data Extraction Process}\label{sec:data_extraction_process}

Data extraction was carried out independently by four coders. Initially, we conducted a pilot test with all four coders to code 3 randomly selected articles to verify and refine our extraction protocol. This pilot phase helped clarify variable definitions and standardize the extraction approach among coders. The coding book included a total of 35 variables that were required to be coded. Out of the 35 variables, 25 were categorical variables with structured categories, while 10 were open-ended questions. For the 25 categorical variables, the choices were included in the coding book, and the coders selected one/more than one option from them. For the 10 open-ended questions, two researchers encoded them into different sub-categories and categories according to different RQs.
The four coders (post-graduate HCI students) who performed the primary extraction have background knowledge in personas and technical expertise in computing and AI (see Table A4 in the online appendix). Each coder was assigned a subset of articles (three coders analyzed 22 articles each, while one coded 42 articles). All four coders independently analyzed 9 (11.11\% of the corpus) common articles from the corpus to assess inter-rater reliability.
Inter-rater reliability was assessed through 
data for 17 coded variables (except open-ended questions and basic information) among four coders
using Fleiss' Kappa, which indicated a strong agreement ($\kappa$ = 0.86).

The information extraction process involved reading each article and coding the 35 variables according to the predefined protocol in the coding book. Each coder was given the standardized extraction protocol with the coding book, with the option of discussions to resolve any ambiguities. 
During the extraction process, 12 articles were further removed from the corpus, bringing the total number for analysis to 81. Eleven articles were removed because of their unconventional structure and brevity, which resulted in many missing key attributes. One article was excluded due to doubts about its methodological rigor. 

\section{Findings}

\subsection{RQ1: How Are GenAI Technologies Used in Persona Development?}

\subsubsection{Technology Adoption and Usage Patterns} \label{sec:tech_adoption}

Our data show that OpenAI's GPT models appear in 70 of 81 articles (86.4\%), establishing GPT as the predominant choice. 
Claude (Anthropic) and Llama (Meta) each appear in 7 articles (8.6\%), while Gemini (Google) is used in 6 articles (7.4\%). The ``Others'' category includes 19 articles (23.5\%) using models such as Deepseek, Mixtral, Bard, and GLM. 
Most of the articles employ only one LLM (n = 53, 65.4\%), while 20 articles (24.7\%) use multiple LLMs together. The remaining 8 articles (9.9\%) do not mention their LLM usage.

The reason for using multiple LLMs tends to be their comparison using quantitative metrics.
For example, Xu et al. \cite{xu_character_2024} compared Claude-3.5, Gemini-1.5, and Llama-3 for persona generation across different metrics, and Sethi et al. \cite{sethi_when_2025} tested GPT-4, Gemini, and Claude for lexical diversity analysis.

The integration of conventional ML approaches with GenAI methods occurs in a relatively few (n = 14, 17.3\%) articles. 
Notable examples of this hybrid approach include using NLP to extract entities from textual data for prompt enrichment \cite{zhang_auto-generated_2024}, and using clustering techniques to segment embeddings before persona generation \cite{li_consumer_2025}.
A key takeaway is \textbf{OpenAI's GPT models overwhelmingly prevail GenAI persona research, creating a possible dependency of personas on this one commercial actor and accentuating limited methodological diversity.}

\subsubsection{Usage Across the Persona Development Lifecycle}

We used a taxonomy adopted from persona literature \cite{tseng_two_2024,pruitt2010persona} to organize GenAI usage patterns into four themes (see Table~\ref{tab:usage_pattern}): (1) persona creation and enrichment, (2) persona-based simulation and interaction, (3) persona-informed personalization, and (4) validation and evaluation.

\begin{figure*}[ht]
    \centering

    \begin{subfigure}[c]{0.65\textwidth}
        \centering
        \includegraphics[width=\linewidth]{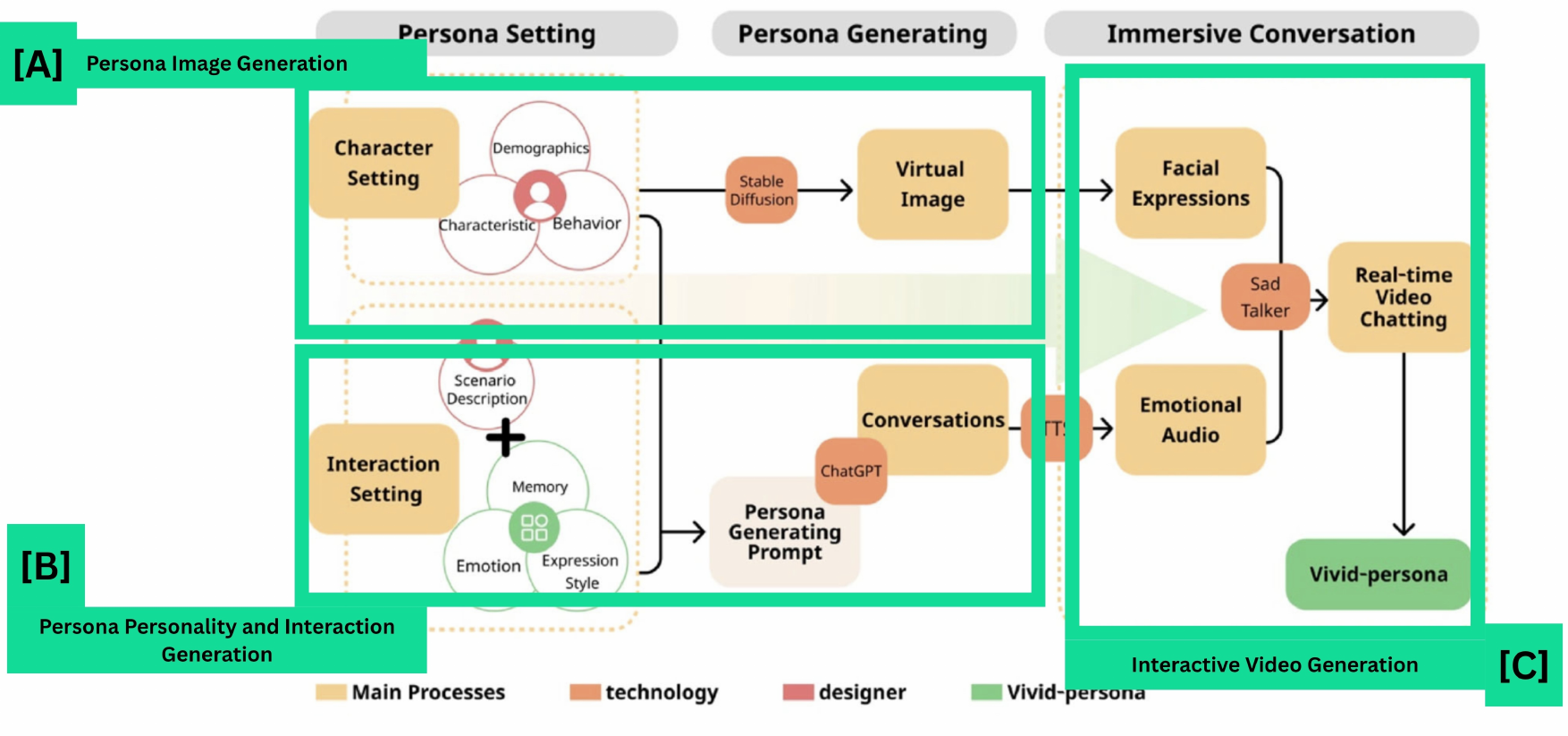}
        \caption{A multimodal persona workflow}
        \label{fig:multimodal_workflow}
    \end{subfigure}
    \hfill
    \begin{subfigure}[c]{0.34\textwidth}
        \centering
        \includegraphics[width=0.8\linewidth]{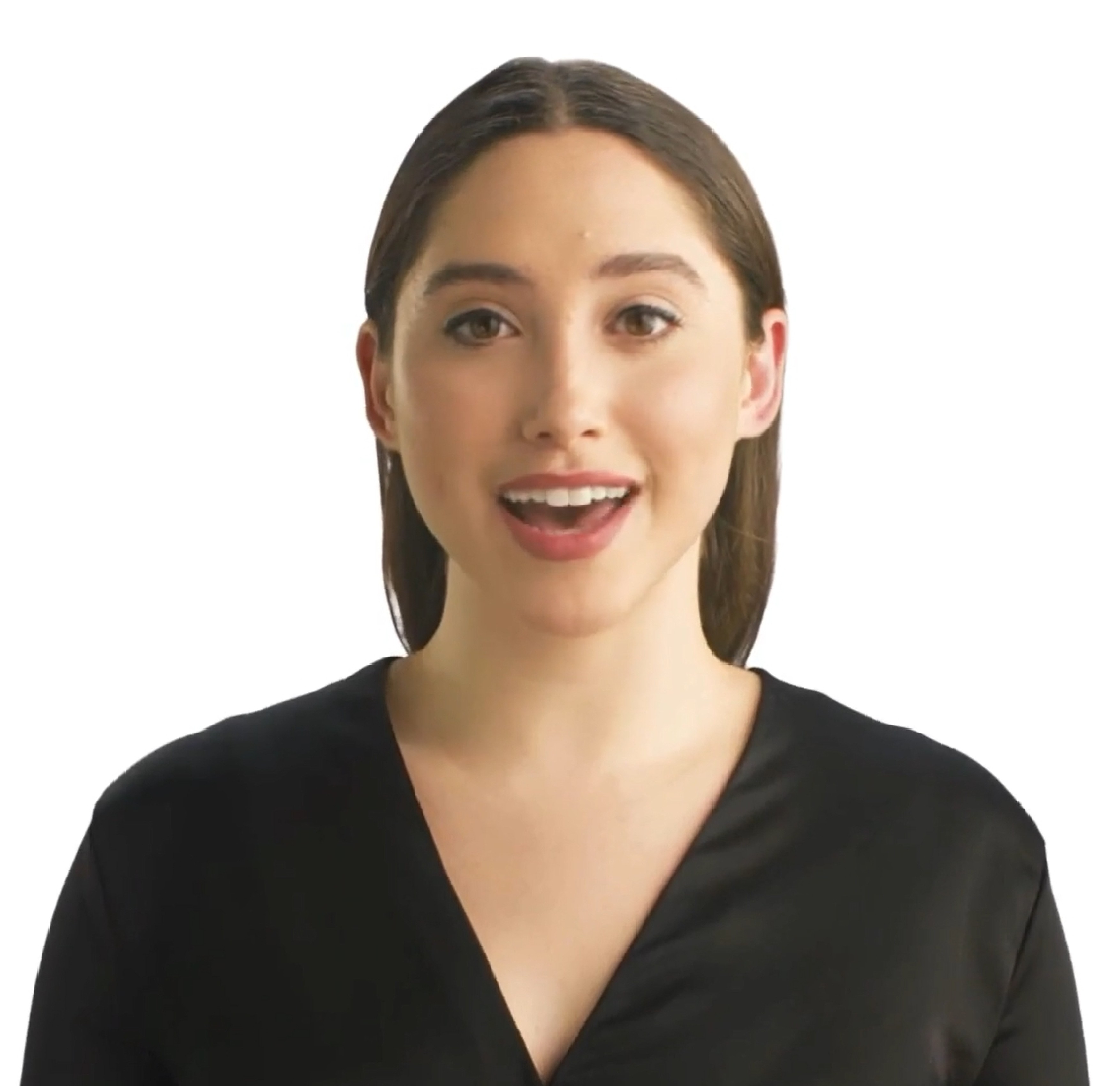}
        \caption{Deepfake personas}
        \label{fig:deepfake_persona}
    \end{subfigure}   
    
    \caption{\textbf{(a)} Vivid-Persona system architecture \cite{zhou_vivid-persona_2024} with annotated phases: Persona Setting ([A]), Persona Generating ([B]), and Immersive Conversation ([C]), combining Stable Diffusion, ChatGPT, and real-time video generation for interactive persona creation. \textbf{(b)} A deepfake persona created using GenAI, illustrating how ``harmful'' technology \cite{kaate_there_2024,hynek_risks_2025} can be repurposed for positive user representation. A video version of the deepfake persona is shared with the author's permission in our online supplementary material\protect\footnotemark{} to further illustrate the interactive nature of some GenAI persona types. The figure is annotated to explain different components of a GenAI persona development system.}
\label{fig:multimodal}
\Description{Two panels showing advanced persona generation techniques. Panel A displays the Vivid-Persona system architecture with three annotated phases: Persona Setting in [A], Persona Generating in [B], and Immersive Conversation in [C], combining Stable Diffusion, ChatGPT, and real-time video generation. Panel B shows a deepfake persona created using GenAI, illustrating how technology traditionally considered harmful can be repurposed for positive user representation.}
\end{figure*}

\textit{Persona Creation and Enrichment:}
We found three main approaches to using GenAI in persona development. First, \textit{data-driven persona generation} creates personas from structured datasets (e.g., Likert-based surveys \cite{jung2025personacraft} and using grouped consumer data \cite{li_consumer_2025}). Second, \textit{narrative persona profiles and enrichment} create detailed storytelling personas with rich backstories \cite{schuller_generating_2024,bai_agentic_2024}. Third, \textit{multimodal persona representation} combines textual descriptions with visual elements \cite{zhang_auto-generated_2024,voigt_re-imagen_2024}. Figure \ref{fig:multimodal} shows an example workflow for multimodal personas.

Participatory approaches appear in 30.9\% of articles (n = 25), while 65.4\% (n = 53) do not involve direct user participation. The remaining 3 articles do not explicitly mention their use of participatory design. Here, participatory persona development involves direct user feedback from either (a) the persona users or (b) the people represented by the personas to modify the personas iteratively through feedback sessions \cite{sun_persona-l_2025}. 



\footnotetext{\url{https://osf.io/3pwav?view_only=29c0a9a1251b43a186e1359ac9cd6a10}}

\begin{figure}
    \centering
    \includegraphics[width=\linewidth]{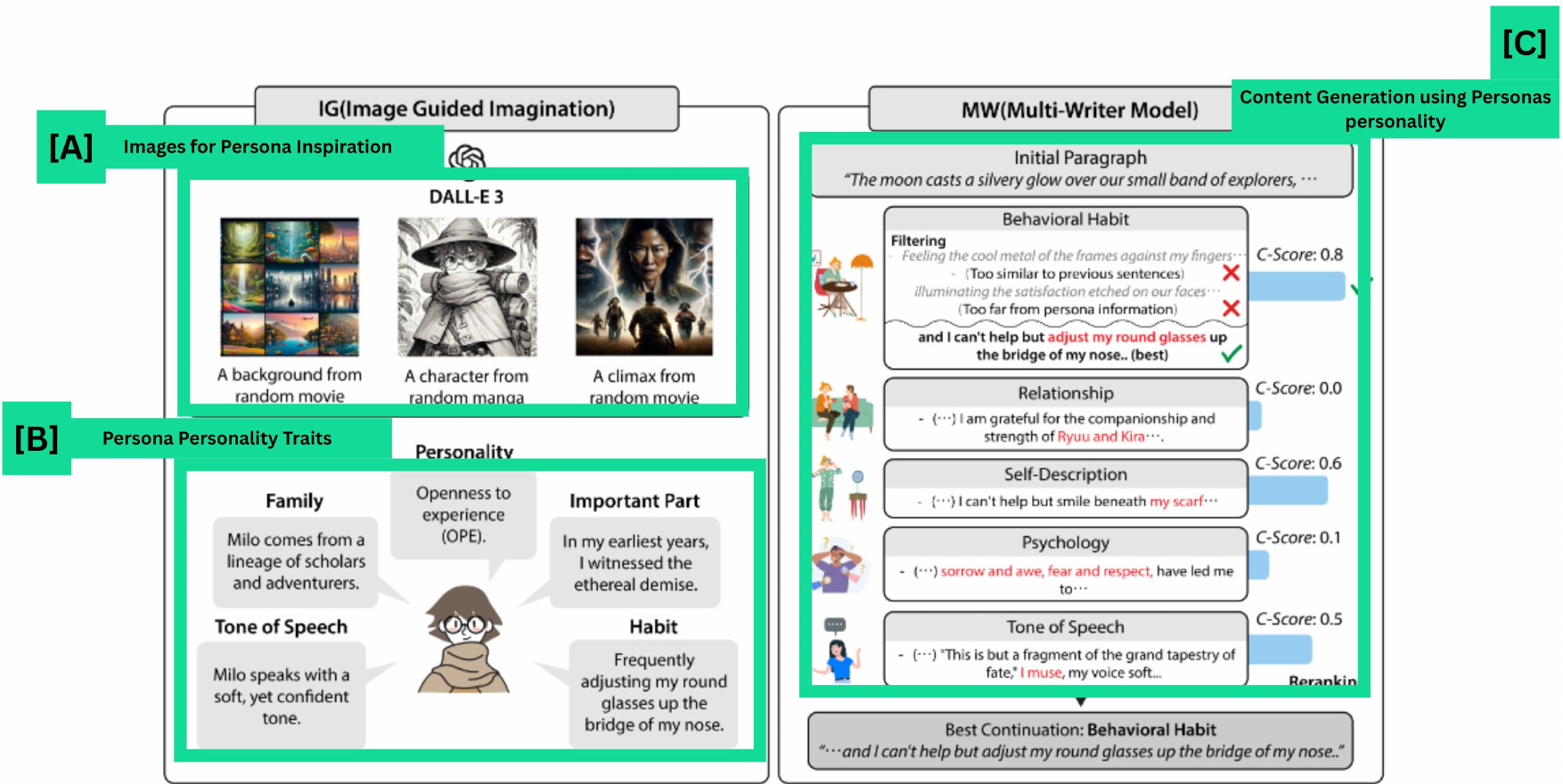}
    \caption{A framework using images for persona generation \cite{park_character-centric_2024} with two annotated components: Image Guided Imagination (IG) using DALL-E 3 ([A]) and Persona Personality Traits ([B]). The Multi-Writer Model([C]) generates content for specific personas, illustrating how GenAI personas can be embedded to automate decision-making in user-centric tasks. The figure is annotated to explain different components of a GenAI persona development system using different data sources.}
    \Description{A framework for image-guided persona generation with two main components. The [A] section shows Image Guided Imagination using DALL-E 3, and the [B] section displays Persona Personality Traits. The Multi-Writer Model generates content for specific personas, demonstrating how AI-based personas can be embedded to automate decision-making in user-centric tasks.}
    \label{fig:character}
\end{figure}

\textit{Persona-based Simulation and Interaction:}
One novelty of GenAI is the ability of LLMs to orient personas in interactive scenarios, which can be implemented in different ways. For example, \textit{conversational persona agents} enable direct user interaction with personas \cite{lo_noel_2025,alqadi2025ramadanpersonas}, such as using LLMs for character imagination and story generation in a dialogue environment \cite{park_character-centric_2024} (see Figure \ref{fig:character}). Another example is using \textit{multi-agent persona simulation}, which involves multiple personas interacting simultaneously; for example, role-playing language agents in decision-making scenarios \cite{dong_can_2024} and deploying multi-agent persona simulation systems for training purposes \cite{rudolph_ai-based_2024}.

\textit{Persona-informed Personalization:}
This application adapts LLM outputs based on persona characteristics. \textit{Personalized content and recommendations} include articles exploring how personas can guide content generation and recommendation systems. \textit{Adaptive dialogue and instruction} adjust communication styles, as exemplified by the \textit{Persona-L} system that developed adaptive dialogue capabilities to represent people with complex needs, allowing the system to adjust its communication style according to persona characteristics \cite{sun_persona-l_2025}.

\begin{figure}
    \centering
    \includegraphics[width=0.75\linewidth]{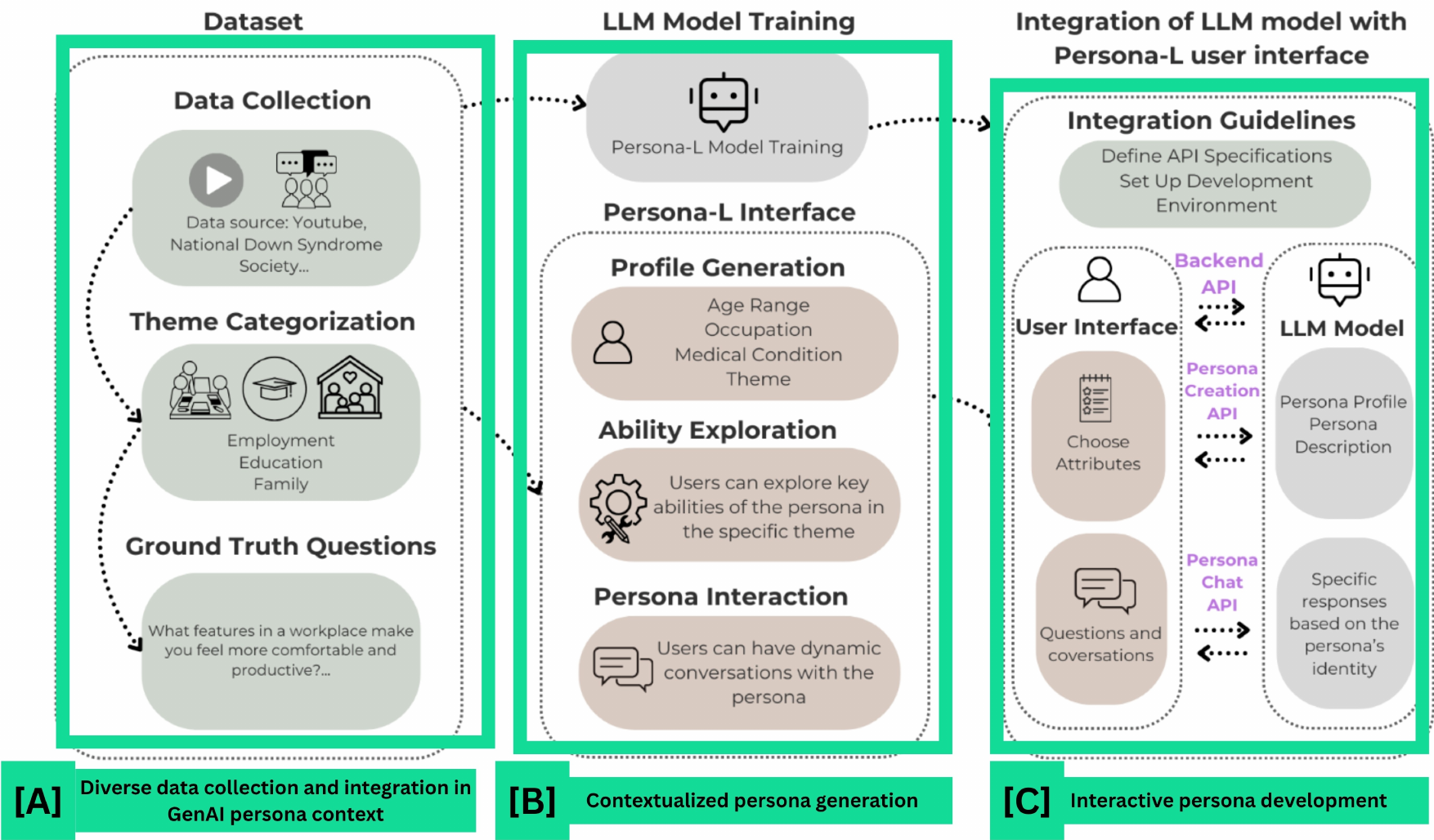}
    \caption{Persona-L framework \cite{sun_persona-l_2025} showing three annotated phases: Dataset collection and integration ([A]); LLM Model Training ([B]) featuring interface with profile generation, ability exploration, and persona interaction capabilities; and Integration with user interface ([C]) providing backend API connections for persona creation and chat functionality. Persona-L illustrates how GenAI models can contribute to varying tasks in persona creation, validation, and delivery using multimodal data sources. The figure is annotated to explain different components of a Persona-L system.}
    \label{fig:context}
    \Description{The Persona-L framework showing three phases with coded annotations. The [A] section covers dataset collection and integration, the [B] section features LLM Model Training with profile generation and persona interaction capabilities, and the [C] section shows integration with a user interface providing backend API connections for persona creation and chat functionality.}
\end{figure}

The use of LLM with personas involves the opportunity for HCI researchers to communicate with personas, rather than passively scanning persona profiles. An example workflow for an adaptive persona is visible in Figure \ref{fig:context}.

\textit{Validation and Evaluation:}
The articles apply LLMs in different ways to assess the quality and consistency of personas. First, \textit{consistency checking} examines whether personas maintain coherent characteristics across scenarios, such as using consistency measurements to validate personas' textual coherence \cite{sun_building_2024}, and developing evaluation metrics such as fluency and helpfulness to assess persona quality \cite{li_iqa-eval_2024}. Second, \textit{realism assessment} evaluates human-likeness through metrics such as lexical diversity \cite{sethi_when_2025} and understanding hallucinations in LLM-generated personas \cite{kaate_you_2025}. Third, \textit{bias detection} identifies stereotypical representations, as exemplified by Panda et al. \cite{panda_llms_2024} who investigated whether GenAI personas perpetuate harmful stereotypes.


\begin{table*}[ht]
{\footnotesize
\caption{LLM usage in persona development with examples and specific applications. While LLMs are primarily used in persona creation and enrichment stages, other modes that use LLMs are simulation, interaction, personalization, validation, and evaluation. See Table A5 in the online appendix for a complete list of articles.}
\label{tab:usage_pattern}

\begin{tabular}{p{2.2cm} p{2.8cm} p{2cm} p{6.8cm}}
\toprule
\cell{\textbf{Main theme}} & \cell{\textbf{Sub-theme}} & \cell{\textbf{Examples}} & \cell{\textbf{Specific application example}} \\
\midrule

\multirow{9}{=}{\cell{\textbf{Persona creation and enrichment}}}
& \cell{Data-driven persona generation}
& \cell{\cite{jung2025personacraft,li_consumer_2025,shin_understanding_2024}}
& \cell{PersonaCraft processes survey responses to automatically generate personas with demographic profiles and segmented information from the survey} \\
\cmidrule(lr){2-4}

& \cell{Narrative persona profiles and enrichment}
& \cell{\cite{schuller_generating_2024,bai_agentic_2024,paoli_writing_2023}}
& \cell{Creating rich storytelling personas with detailed backstories, personality descriptions, and contextual narratives for design scenarios} \\
\cmidrule(lr){2-4}

& \cell{Multimodal persona representation}
& \cell{\cite{zhang_auto-generated_2024,voigt_re-imagen_2024,chen_empathy-based_2024}}
& \cell{Generating personas that combine textual descriptions with visual avatars, mood boards, and lifestyle imagery} \\
\midrule

\multirow{6}{=}{\cell{\textbf{Persona-based simulation and interaction}}}
& \cell{Conversational persona agents}
& \cell{\cite{park_character-centric_2024,lo_noel_2025,choi_proxona_2025}}
& \cell{Developing interactive chatbots that embody specific personas, allowing users to have conversations with target user archetypes} \\
\cmidrule(lr){2-4}

& \cell{Multi-agent persona simulation}
& \cell{\cite{dong_can_2024,rudolph_ai-based_2024,xu_character_2024}}
& \cell{Simulating group discussions between multiple persona agents to understand consensus-building and decision-making processes} \\
\midrule

\multirow{6}{=}{\cell{\textbf{Persona-informed personalization}}}
& \cell{Personalized content and recommendations}
& \cell{\cite{li_consumer_2025,mondal_presentations_2024}}
& \cell{Tailoring marketing messages and product recommendations based on specific persona characteristics and preferences} \\
\cmidrule(lr){2-4}

& \cell{Adaptive dialogue and instruction}
& \cell{\cite{sun_persona-l_2025,kaiser_simulating_2025}}
& \cell{Adjusting communication style and accessibility features based on users' cognitive abilities and interaction preferences} \\
\midrule

\multirow{12}{=}{\cell{\textbf{Validation and evaluation}}}
& \cell{Consistency checking}
& \cell{\cite{sun_building_2024,li_iqa-eval_2024,kaate_fourth_2025}}
& \cell{Evaluating whether persona responses remain consistent across different scenarios and interaction contexts} \\
\cmidrule(lr){2-4}

& \cell{Realism assessment}
& \cell{\cite{sethi_when_2025,smrke_exploring_2025}}
& \cell{Measuring linguistic patterns to assess how human-like generated personas appear} \\
\cmidrule(lr){2-4}

& \cell{User preference prediction}
& \cell{\cite{yeykelis_using_2024,kaiser_simulating_2025}}
& \cell{Testing whether LLM-generated personas can accurately predict real user preferences in product choices and behavioral decisions} \\
\cmidrule(lr){2-4}

& \cell{Bias detection}
& \cell{\cite{panda_llms_2024,park_as_2025}}
& \cell{Analyzing generated personas for demographic stereotypes, cultural biases, and unfair generalizations across different user groups} \\
\bottomrule
\end{tabular}
}
\end{table*}

\subsubsection{Data Sources and Representation Approaches}  \label{sec:data_sources}

Data collection practices are split among real user data, which appears in approximately half (n = 41, 50.6\%) of the articles, synthetic data in 32.1\% (n = 26), and mixed approaches in 17.3\% (n = 14). 
Researchers used real data collected directly from the users through surveys \cite{jung2025personacraft, li_consumer_2025} or indirectly collected through observations \cite{gupta_evaluation_2024}. Synthetic data applications, for example, used artificially generated data to generate personas representing addiction patterns \cite{sethi_when_2025} and investigated the consistency of LLM-generated personas using personality tests  \cite{de_winter_use_2024}. For example, Bai et al. \cite{bai_agentic_2024} used mixed methods combining real and synthetic data to create virtual populations by sampling census-based skeletal personas and enriching them with LLM-generated details.

\begin{figure}
    \centering
    \includegraphics[width=\linewidth]{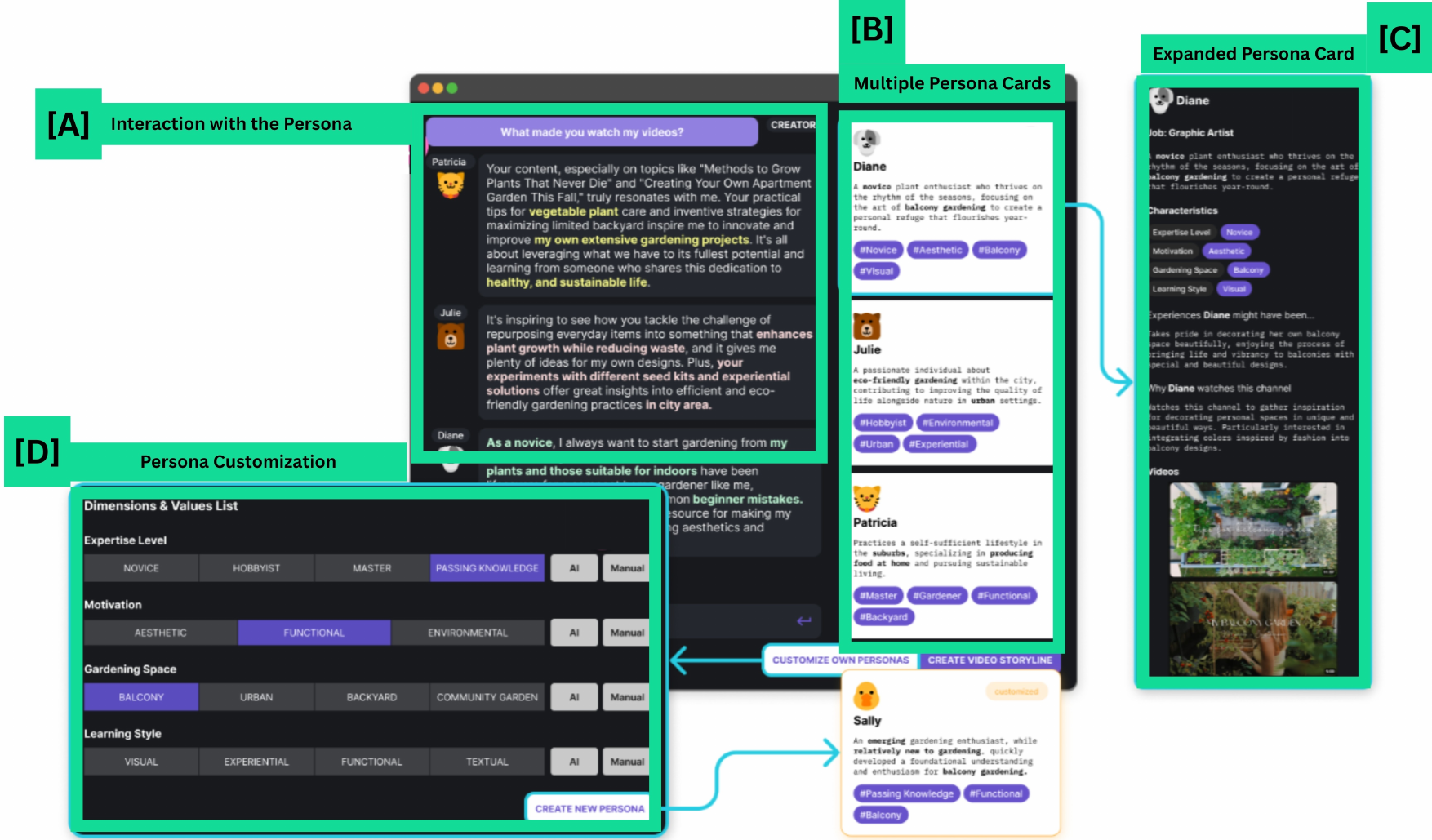}
    \caption{Proxona, a multiple persona interactive tool \cite{choi_proxona_2025} with annotated sections: Persona Customization ([D]), Multiple Persona Cards ([B]), and Expanded Persona Card (c). The tool demonstrates editing and interaction capabilities ([A]) with multiple personas, illustrating the complexities of creating full-fledged persona systems that rely on multiple system prompts using LLMs for persona creation, evaluation, and delivery.The figure is annotated to explain different components of a Proxona system.}
\label{fig:proxona}
\Description{ The Proxona multiple persona interactive tool with four annotated sections. The [D] section shows Persona Customization controls, the [B] section displays Multiple Persona Cards in a grid layout, and the [C] section presents an Expanded Persona Card with detailed information. The tool demonstrates editing and interaction capabilities [A] with multiple personas.}
\end{figure}

Persona representation formats reveal both traditional and emerging patterns. 
Traditional persona profiles are the dominant form of presenting personas (n = 53, 65.4\%), providing structured demographic and behavioral information in standard persona templates \cite{markey2025framework,li_actions_2025,mercer2025applying}. In contrast, narrative formats appear in approximately a third of the articles (n = 26, 32.1\%); for example, De Paoli \cite{paoli_writing_2023} produced persona narratives using LLM-assisted thematic analysis, while Schuller et al. \cite{schuller_generating_2024} generated persona narratives as user stories that UX experts found comparable with human-written personas. Furthermore, personas are presented as chatbots in 21.0\% of articles (n = 17), typically in interactive persona interfaces \cite{sun_persona-l_2025,choi_proxona_2025} that combine profile and dialogue formats for the persona user (see Figure \ref{fig:proxona}). Finally, another notable development is providing persona information in structured data formats (n = 5, 6.2\%), for example, using a nested JSON output that makes it straightforward to quantitatively analyze the persona attributes \cite{salminen_deus_2024}.

\subsubsection{Application Contexts}
GenAI personas are applied across diverse domains and contexts (see Table A6 in the online appendix for a complete list of articles). Product development represents the most common application context (n = 9, 11.1\%), in which researchers create data-driven personas to improve technology design and UX. Marketing contexts (n = 7, 8.6\%) employ LLMs for consumer segmentation through text embeddings and clustering \cite{li_consumer_2025}, and create demographic profiles that mirror real participants to generate synthetic survey responses \cite{kaiser_simulating_2025}, among other uses.

Education contexts (n = 6, 7.4\%) include, for example, creating chatbot personas representing children with disabilities to support accessible technology design \cite{lo_noel_2025}, whereas healthcare contexts (n = 5, 6.2\%), for example, compare LLM-generated personas with human-crafted personas in clinical research \cite{smrke_exploring_2025} and represent people with special conditions like Down syndrome \cite{sun_persona-l_2025}.
In articles focused on civic good and social applications (n = 7, 8.6\%), researchers analyzed GenAI personas in societally meaningful contexts, such as addiction research \cite{salminen_deus_2024,sethi_when_2025} and investigated systematic biases in GenAI systems \cite{panda_llms_2024}. Research enhancement contexts, in turn, represent 3.7\% (n = 3) of applications, in which researchers explain persona creation processes for novice designers through collaboration with GenAI tools \cite{jung2025personacraft} and use expert perspectives to improve research ideation \cite{liu_personaflow_2025}.

The remaining 54.3\% (n = 44) of applications span diverse contexts, including sustainability, finance, content creation, accessibility design, and language processing, indicating that GenAI personas are being adopted across more or less every domain that traditionally employs personas.

\subsubsection{Alternative Development Approaches}

GenAI personas incorporate at least four novelties or ``innovations'' (see Table~\ref{tab:p4-thematic-coding}): (1) input source innovations, (2) process innovations, (3) prompt engineering innovations, and (4) output format innovations. We explain these innovations as follows.

Input source innovations (n = 26, 32.1\%) expand the types of data used for persona generation beyond traditional surveys and analytics. \textit{Visual and multimodal input} integrates images, avatars, and visual representations into persona profiles, expanding beyond text-only descriptions. Multimodal data improves GenAI personas by combining textual analysis with visual generation. Persona systems can pair GPT-4 survey processing with DALL-E 2 avatar creation \cite{zhang_auto-generated_2024}, or integrate emotional expressions for privacy education \cite{chen_empathy-based_2024}, with the idea that visual elements strengthen memorability beyond text-only profiles. Similarly, unstructured qualitative data extends persona development into contexts where collecting structured, quantitative data is impractical. For example, LLMs can extract personas from video content and industry reports \cite{sun_persona-l_2025} or generate personas from interview transcripts through thematic analysis \cite{paoli_improved_2023}. Expert knowledge can be indirectly used to generate personas; for example, by capturing domain expertise through graph-based simulation \cite{liu_personaflow_2025} or transforming expert knowledge into narrative profiles \cite{kaiser_simulating_2025}. 

Process innovations (n = 30, 37\%) change the fundamental workflow of persona generation, often bypassing steps that were previously considered essential. Data-driven persona development often follows a sequential pipeline of collecting data, identifying segments through clustering, and then enriching profiles with narrative details \cite{salminen2021survey,jung2018automatic}. The use of GenAI challenges this conventional pipeline. 
For example, rather than discovering segments in data, \textit{direct attribute specification} enables researchers to specify desired characteristics upfront and have LLMs generate matching personas. Li et al. \cite{li_consumer_2025} demonstrated this by directly specifying consumer segments to generate personas for predefined market categories, eliminating the need for any preprocessing stages altogether. Similarly, \textit{segmentation-free approaches} that generate personas from data directly using theoretical frameworks without any empirical clustering. Sun et al. \cite{sun_persona-l_2025} used conceptual models and reports to generate personas for people with Down syndrome. \textit{Generation pipelines} represent automating the entire workflow. Systems like PersonaCraft \cite{jung2025personacraft} and PersonaFlow \cite{liu_personaflow_2025} integrate all the stages of persona generation (data collection, analysis, and generation) into unified, automated processes.

Prompt engineering innovations (n = 18, 22.9\%) apply advanced techniques to improve the quality and consistency of LLM-generated personas. \textit{Two-stage prompting} separates the creation of the demographic skeletal profile from the finalization of the persona. For example, researchers first prompt demographic profile creation, followed by enrichment prompts to add behavioral and contextual information \cite{salminen_deus_2024,sethi_when_2025}. Taking this idea further, \textit{multistage prompts} are applied in different steps of the persona generation pipeline. For example, Panda et al. \cite{panda_llms_2024} used sequential prompts for data analysis, persona attribute extraction, and profile writing. Articles also apply \textit{few-shot-learning prompting approaches} that include example personas in the prompt to guide LLM output style and structure. For example, Schuller et al. \cite{schuller_generating_2024} included sample personas in prompts for bias evaluation, helping LLMs understand expected output format and detail level, while Chen et al. \cite{chen_empathy-based_2024} used example privacy scenarios to generate contextually adapted personas.

\label{narrative-output}
Output format innovations (n = 39, 48.1\%) change how personas are presented and interacted with, moving beyond static profile documents. \textit{Narrative-driven personas} present information in a story format rather than attribute lists. Xu et al. \cite{xu_character_2024} created personas with rich backstories and character arcs for creative applications, while Schuller et al. \cite{schuller_generating_2024} generated first-person narratives that read like biographical sketches and can make personas more engaging and memorable for stakeholders. \textit{Multimodal representations} extend this engagement by combining text with visual and emotional elements. Chen et al. \cite{chen_empathy-based_2024} integrated emotional expressions with visual representations to create empathy-based personas, while Zhang et al. \cite{zhang_auto-generated_2024} combined textual personas with auto-generated avatar pictures and lifestyle imagery. \textit{Interactive persona systems} complete this transformation by presenting personas through conversational interfaces rather than static documents. For example, Persona-L \cite{sun_persona-l_2025} provides conversational interfaces for accessibility needs, enabling persona users to discuss requirements with persona representations.

\begin{table*}[htbp]
\centering
\small
\caption{Thematic coding for unique persona development approaches. LLMs bring four innovations to persona generation, including the ability to include different data sources, change the process of generating personas, prompting innovations, and innovations regarding interaction with personas. Articles are assigned to one primary sub-theme per main category and may appear in more than one main category.}
\label{tab:p4-thematic-coding}
\begin{tabular}{p{3cm}p{3cm}p{8cm}}
\toprule
\textbf{Main theme} & \textbf{Sub-theme} & \textbf{Papers (n)} \\
\midrule

\multirow{3}{*}{\parbox[t]{3cm}{\raggedright\textbf{Input source innovations} (n = 25, 31.2\%)\par}} 
& Visual/multimodal input 
& \cite{fujimoto_ghost_2025}, \cite{park_character-centric_2024}, \cite{salminen_picturing_2024}, \cite{sun_kiss_2024}, \cite{zhang_auto-generated_2024}, \cite{zhou_vivid-persona_2024} (n = 6) \\
\cmidrule(lr){2-3}

& Alternative data sources 
& \cite{clements_innovative_2023}, \cite{farseev_somonitor_2024}, \cite{goel_preparing_2023}, \cite{gupta_evaluation_2024}, \cite{kaate_fourth_2025}, \cite{kaate_you_2025}, \cite{oksman_drone_2025}, \cite{paoli_improved_2023}, \cite{paoli_writing_2023}, \cite{rudolph_ai-based_2024}, \cite{schirgi_development_2024}, \cite{yin2025co} (n = 12) \\
\cmidrule(lr){2-3}

& Expert-generated content 
& \cite{benharrak_writer-defined_2024}, \cite{doudkin_synthetic_2025}, \cite{huang_exploring_2024}, \cite{liu_personaflow_2025}, \cite{nobrega2025towards}, \cite{sun_persona-l_2025}, \cite{yin2025co}, \cite{zheng_customizing_2025} (n = 8) \\
\midrule

\multirow{3}{*}{\parbox[t]{3cm}{\raggedright\textbf{Process innovations} (n = 30, 37.5\%)\par}} 
& Direct attribute specification 
& \cite{li_actions_2025}, \cite{li_chatgpt_2024}, \cite{li_consumer_2025}, \cite{luo2024sociotechnical}, \cite{mayr2025chatchecker}, \cite{park_as_2025}, \cite{rudolph_ai-based_2024}, \cite{venkit_tale_2025}, \cite{zhou_vivid-persona_2024} (n = 9) \\
\cmidrule(lr){2-3}

& Segmentation-free approaches 
& \cite{park_social_2022}, \cite{schuller_generating_2024}, \cite{sun_persona-l_2025} (n = 3) \\
\cmidrule(lr){2-3}

& Novel generation pipelines 
& \cite{arora2025ai}, \cite{bai_agentic_2024}, \cite{chen_empathy-based_2024}, \cite{clements_innovative_2023}, \cite{gonzalez_exploring_2024}, \cite{jung2025personacraft}, \cite{karolita_crafter_2024}, \cite{markey2025framework}, \cite{paoli_improved_2023}, \cite{paoli_writing_2023}, \cite{park_audilens_2023}, \cite{park_character-centric_2024}, \cite{salminen_deus_2024}, \cite{shin_understanding_2024}, \cite{voigt_re-imagen_2024}, \cite{wongabut_automating_2025}, \cite{wu2025aligning}, \cite{zhang_auto-generated_2024} (n = 18) \\
\midrule

\multirow{4}{*}{\parbox[t]{3cm}{\raggedright\textbf{Prompt engineering innovations} (n = 18, 22.5\%)\par}} 
& Two-stage prompts 
& \cite{gonzalez_exploring_2024}, \cite{salminen_deus_2024}, \cite{sethi_when_2025} (n = 3) \\
\cmidrule(lr){2-3}

& Multistage prompts 
& \cite{paoli_improved_2023}, \cite{shin_understanding_2024} (n = 2) \\
\cmidrule(lr){2-3}

& Chain-of-thought methods 
& \cite{park_character-centric_2024}, \cite{wongabut_automating_2025} (n = 2) \\
\cmidrule(lr){2-3}

& Few-shot learning 
& \cite{arora2025ai}, \cite{barambones_chatgpt_2024}, \cite{chen_empathy-based_2024}, \cite{cheng_marked_2023}, \cite{dong_can_2024}, \cite{li_actions_2025}, \cite{li_chatgpt_2024}, \cite{maharana_evaluating_2024}, \cite{mercer2025applying}, \cite{schuller_generating_2024}, \cite{wu2025aligning} (n = 11) \\
\midrule

\multirow{3}{*}{\parbox[t]{3cm}{\raggedright\textbf{Output format innovations} (n = 44, 55.0\%)\par}} 
& Narrative-driven personas 
& \cite{bai_agentic_2024}, \cite{chen_why_2024}, \cite{cheng_marked_2023}, \cite{goel_preparing_2023}, \cite{gonzalez_exploring_2024}, \cite{mercer2025applying}, \cite{oksman_drone_2025}, \cite{paoli_writing_2023}, \cite{park_character-centric_2024}, \cite{park_social_2022}, \cite{salminen_deus_2024}, \cite{schuller_generating_2024}, \cite{sethi_when_2025}, \cite{smrke_exploring_2025}, \cite{venkit_tale_2025}, \cite{wongabut_automating_2025} (n = 16) \\
\cmidrule(lr){2-3}

& Multimodal representations 
& \cite{chen_empathy-based_2024}, \cite{fujimoto_ghost_2025}, \cite{morande2023digital}, \cite{park_character-centric_2024}, \cite{salminen_picturing_2024}, \cite{voigt_re-imagen_2024}, \cite{zhang_auto-generated_2024}, \cite{zhou_vivid-persona_2024} (n = 8) \\
\cmidrule(lr){2-3}

& Interactive persona systems 
& \cite{barambones_chatgpt_2024}, \cite{chen2025recusersim}, \cite{chen2025textsc}, \cite{choi_proxona_2025}, \cite{doudkin_synthetic_2025}, \cite{he2025simupanel}, \cite{huang_exploring_2024}, \cite{kaate_you_2025}, \cite{kaiser_simulating_2025}, \cite{kaur_synthetic_2025}, \cite{li_consumer_2025}, \cite{lo_noel_2025}, \cite{maharana_evaluating_2024}, \cite{milovic2024sell}, \cite{nguyen_simulating_2024}, \cite{nobrega2025towards}, \cite{park_audilens_2023}, \cite{rudolph_ai-based_2024}, \cite{sun_persona-l_2025}, \cite{yates_using_2024}, \cite{zheng_customizing_2025} (n = 21) \\
\bottomrule

\end{tabular}
\end{table*}

\subsection{RQ2: How Are GenAI Personas Evaluated?}

\subsubsection{Evaluation Targets in GenAI Personas}

We identified four primary aspects evaluated in GenAI-generated persona research: (1) persona descriptions, (2) persona behavior, (3) generation process, and (4) outcomes and impact (Table~\ref{tab:what_evaluated}). Among articles that reported evaluation (n = 44), description quality was examined in 38.6\% of articles, behavioral performance in 36.4\%, outcomes/impact in 15.9\%, and generation process in 11.4\%.

\textit{Persona Description Evaluation:}
Description evaluation examines the static attributes and textual quality of persona profiles. Articles assess whether persona descriptions appear credible, internally consistent, and linguistically sound. Generally, this is done by using established HCI scales such as the Persona Perception Scale (PPS) \cite{salminen_persona_2020}. The PPS is an instrument for evaluating individuals' perceptions of personas, comprising 8 measures such as credibility, completeness, consistency, clarity, 
and willingness to use \cite{salminen_persona_2020}. Multiple articles applied PPS to evaluate description in GenAI personas \cite{smrke_exploring_2025,kaate_fourth_2025}. For example, Jung et al. \cite{jung2025personacraft} evaluated personas on five dimensions: consistency, clarity, completeness, credibility, and fluency.
Computational approaches to description evaluation include linguistic analysis to measure lexical diversity \cite{sethi_when_2025} and stereotype detection using predefined lexicons \cite{cheng_marked_2023}. These methods assess whether persona descriptions employ varied vocabulary and whether they contain problematic stereotypical representations.

\textit{Persona Behavior Evaluation:}
Behavior evaluation examines personas' performance in interactive contexts or when tasked with decision-making, prediction, or simulation activities. This aspect differs from description evaluation: description evaluation asks whether a persona profile appears credible, and behavior evaluation asks whether the persona acts in ways that match real human patterns.
Articles employing behavioral evaluation tested personas in varied contexts. For example, Li et al. \cite{li_consumer_2025} validated GenAI personas against actual survey data, demonstrating that GenAI personas can simulate consumer preferences. Kaiser et al. \cite{kaiser_simulating_2025} conducted pre- and post-intervention surveys measuring belief changes and information sharing patterns among real humans, simulated humans, and synthetic personas. Park et al. \cite{park_character-centric_2024} tested the contextual appropriateness of personas' responses in naturalistic decision-making scenarios. 
However, behavioral evaluation remains less developed than description evaluation. Articles often assess behavior through indirect measures rather than testing of persona actions in controlled scenarios.

\textit{Generation Process Evaluation:}
A small subset of articles evaluated the efficiency, scalability, or usability of the persona generation process itself. Zhang et al. \cite{zhang_auto-generated_2024} developed custom metrics measuring generation efficiency, satisfaction with the creation process, and collaboration quality between humans and GenAI systems. These evaluations address practical concerns about integrating GenAI personas into existing decision-making workflows rather than assessing the personas themselves.

\textit{Outcomes and Impact Evaluation:}
Outcome evaluation examines the downstream effects of using GenAI personas in design, decision-making, or educational contexts. For example, articles measured usability using the System Usability Scale (SUS) \cite{kaate_fourth_2025}, assessed learning gains from persona-based educational interventions, and evaluated whether personas influenced design decisions appropriately. This evaluation category connects persona quality to real-world applications but remains relatively rare in the literature.

\begin{table*}[htbp]
\centering
\small 
\caption{Target aspects of GenAI personas that are evaluated. Description and behavioral evaluations receive similar attention, while process and outcome evaluations are less common. See Table A7 in the online appendix for a complete list of articles.}
\label{tab:what_evaluated}
\begin{tabular}{p{3.5cm}p{6cm}p{4cm}}
\toprule
\textbf{Evaluation Aspect} & \textbf{What Is Assessed} & \textbf{Example Articles} \\
\midrule
\textbf{Persona Description} (n = 17, 38.6\%) 
& Static attributes: consistency, clarity, credibility, completeness, fluency, linguistic quality, stereotype presence 
& \cite{jung2025personacraft,cheng_marked_2023,sethi_when_2025} \\
\midrule
\textbf{Persona Behavior} (n = 16, 36.4\%) 
& Dynamic performance: decision-making accuracy, response patterns, interaction quality, alignment with real human behavior 
& \cite{li_consumer_2025,kaiser_simulating_2025,park_character-centric_2024} \\
\midrule
\textbf{Outcomes \& Impact} (n = 7, 15.9\%) 
& Usability, user satisfaction, learning gains, effectiveness in design contexts 
& \cite{kaate_fourth_2025} \\
\midrule
\textbf{Generation Process} (n = 5, 11.4\%) 
& Efficiency, scalability, human-AI collaboration quality, workflow integration 
& \cite{zhang_auto-generated_2024} \\
\bottomrule
\end{tabular}
\end{table*}

\subsubsection{Evaluation Methodologies} \label{sec:eval_methods}
Among all the articles, 44 articles( 54.32\%) mention the evaluation methodology used in the process, while 45.67\%  (n = 37) do not provide a clear evaluation framework.
The use of LLMs in persona evaluation remains limited, with a quarter (n = 12, 27.27\%) of articles employing LLMs for evaluation purposes, while most studies applying evaluation (n = 32, 72.72\%) rely on traditional evaluation techniques. 

We identify three primary evaluation approaches from the articles that mentioned their evaluation approaches. (see Table~\ref{tab:taxonomy_eval_in_AIpersonas}): (1) human-driven evaluation, (2) computational evaluation, and (3) benchmark-based evaluation.

\textit{Human-Driven Evaluation:}
More than half of the articles (n = 25, 58.6\%) that applied the evaluation process used human evaluators to judge persona quality, though it is equally noteworthy that nearly half (41.4\%) did not use human evaluators. For example, Jung et al. \cite{jung2025personacraft} recruited 148 participants, including both regular users (n = 127) and UX experts (n = 21), reporting that GenAI personas scored significantly higher than non-LLM personas for consistency, clarity, completeness, credibility, and fluency. 
Another article measured GenAI personas for authenticity, clarity, empathy, and willingness to use, finding satisfactory quality \cite{zhou_vivid-persona_2024}. Chat-based personas were compared with profile-based personas using both PPS and SUS, with significant differences in usability between interactive and static personas \cite{kaate_fourth_2025}.

Expert analysis represents a valuable approach within human-driven evaluation. Sun et al. \cite{sun_persona-l_2025} used qualitative interviews with domain experts to scope how accurately personas represent people with complex needs. Another article recruited 127 clinical and educational experts to compare LLM-generated personas with human-crafted personas in the obesity research domain \cite{smrke_exploring_2025}.
Discrimination tests provide another human-driven approach. Researchers have used Turing-test style evaluation methods in which a human has to differentiate between human-generated personas and GenAI personas \cite{schuller_generating_2024}. As an example, a two-phase study with 31 participants compared single versus multiple expert perspectives for relevance, helpfulness, and creativity \cite{liu_personaflow_2025}. Another article conducted qualitative interviews to evaluate the effectiveness of chat personas by examining empathy in educational contexts \cite{lo_noel_2025}.

\textit{Computational Evaluation:}
Computational evaluation uses computer algorithms to objectively assess persona quality through structured frameworks and NLP metrics. Around one-fifth (20.7\%) of the articles used this form of persona evaluation. Multiple articles applied computational linguistic tools, such as structured linguistic analysis, sentiment analysis, and stereotype lexicons, to investigate GenAI persona descriptions, for example, detecting potentially harmful stereotypes  \cite{cheng_marked_2023} or investigating how lexically diverse the persona descriptions were \cite{sethi_when_2025}.
A lexical analysis across multiple LLMs (GPT, Gemini, Llama, Claude) found that personas generated by Gemini 1.5 Pro had the highest lexical diversity at the individual persona level, meaning each persona description employed a richer and more varied vocabulary to characterize user attributes \cite{sethi_when_2025}. 
Custom metrics represent another computational approach. For example, articles have developed custom evaluation dimensions that measure efficiency, satisfaction, collaboration, accuracy, creativity, and diversity in persona generation \cite{zhang_auto-generated_2024}. These measures are used for evaluation by the users and are collected via surveys and interviews. Articles \cite{chen_empathy-based_2024} also use user perception scales like the PPS by asking an LLM to evaluate personas according to factors such as clarity, credibility, consistency, and empathy.

\textit{Benchmark-Based Evaluation:}
Benchmark-based evaluation was equally common as the use of computational evaluation (20.7\% of articles).
This approach compares personas with established standards, reference data, or existing frameworks to assess their accuracy and validity. For example, Gupta et al. \cite{gupta_evaluation_2024} evaluated personas representing technology professionals against LinkedIn data, reporting an overrepresentation of elite universities (72.45\% in generated personas vs. 8.56\% in actual data). 
In another article, GenAI personas achieved high accuracy (89\%) in simulating consumer preferences compared to actual survey data, with researchers contending that GenAI personas can accurately represent real consumer behaviors and preferences \cite{li_consumer_2025}. 
Some articles applied benchmarking by comparing LLM-generated personas with human ground truth data using controlled experimental conditions \cite{dong_can_2024} or compared generated personas with theoretical archetypes like ``Suggestible Sally'' using trait alignment analysis \cite{clements_innovative_2023}.
Researchers have also applied pre- and post-intervention surveys to measure established beliefs, behavioral measures, and information sharing among real humans, simulated humans, and synthetic personas to understand the effectiveness of GenAI personas in representing human responses \cite{kaiser_simulating_2025}. De Paoli \cite{paoli_improved_2023} notes that conventional UCD methodologies and established persona development methodologies provide possible baselines for comparison.

\textbf{Human-driven evaluation prevails among the articles. However, a serious concern is that nearly half of all the articles provide no clear evaluation strategy for GenAI personas.}

\begin{table}[htbp]
\centering
\small
\caption{Evaluation methodologies show that GenAI personas are mostly evaluated by humans, but other methods also include evaluating using algorithms and by comparison with established systems/data. See Table A8 in the online appendix for a complete list of articles.}
\label{tab:taxonomy_eval_in_AIpersonas}
\begin{tabular}{p{2.5cm}p{2.5cm}p{1.5cm}}
\toprule
\textbf{Main Theme} & \textbf{Sub-Theme} & \textbf{Example Articles} \\
\midrule

\multirow{4}{3.5cm}{\textbf{Human-Driven \break Evaluation \break (n = 25, 58.6\%)}} 
& Subjective Rating Scales & \cite{jung2025personacraft,zhou_vivid-persona_2024,kaate_fourth_2025} \\
\cmidrule(lr){2-3}
& Expert Analysis & \cite{sun_persona-l_2025,smrke_exploring_2025} \\
\cmidrule(lr){2-3}
& Discrimination Tests & \cite{schuller_generating_2024} \\
\cmidrule(lr){2-3}
& UX Assessment & \cite{liu_personaflow_2025,lo_noel_2025} \\
\midrule

\multirow{2}{3.5cm}{\textbf{Computational \break Evaluation \break (n = 10, 20.7\%)}} 
& NLP & \cite{cheng_marked_2023,sethi_when_2025} \\
\cmidrule(lr){2-3}
& Custom Metrics Frameworks & \cite{zhang_auto-generated_2024,chen_empathy-based_2024} \\
\midrule

\multirow{3}{3.5cm}{\textbf{Benchmark-Based \break Evaluation \break (n = 9, 20.7\%)}} 
& Real-World Data Validation & \cite{gupta_evaluation_2024,li_consumer_2025} \\
\cmidrule(lr){2-3}
& Comparative Standard Analysis & \cite{clements_innovative_2023,paoli_improved_2023} \\
\cmidrule(lr){2-3}
& Mixed-Methods Triangulation & \cite{salminen_deus_2024,kaiser_simulating_2025} \\
\bottomrule
\end{tabular}
\end{table}

\subsubsection{LLMs' Role in Evaluation}

We categorized the LLM use in persona evaluation according to four established validity concepts (Table~\ref{tab:g10-thematic-coding}): (1) face validity, (2) ecological validity, (3) internal validity, and (4) statistical validity. Face validity assesses whether personas appear credible on the surface. Ecological validity examines how personas perform in real-world contexts (e.g., among stakeholders or when integrated into decision support systems). Internal validity evaluates the personas against some technical, predefined criteria (e.g., consistency, diversity), whereas statistical validity relies on statistical testing (estimation, population comparisons).

\textit{Face Validity Assessment:}
In a third of the articles (33.3\%), researchers used LLMs to evaluate whether personas appear credible and realistic on their surface. For example, Li et al. \cite{li_iqa-eval_2024} applied LLMs to assess LLM-generated personas compared to human-annotated examples. Researchers have also used LLMs to evaluate user personas using established questionnaires such as the PPS \cite{panda_llms_2024}, which may provide some form of \textit{face validity} but does not constitute a full assessment of the personas' validity in real decision-making scenarios. 

\textit{Ecological Validity Assessment:}
A quarter of the articles (25.0\%) used LLMs to examine how personas behave in realistic, real-world contexts and scenarios. For example, Park et al. \cite{park_character-centric_2024} tested persona-based decision-making in naturalistic scenarios, with GenAI systems assessing contextually appropriate responses. Researchers used LLMs to evaluate how personas influence guardrail sensitivity based on demographic context \cite{li_chatgpt_2024} and to analyze conversations based on persona attributes in realistic settings \cite{sun_kiss_2024}. Smrke et al. \cite{smrke_exploring_2025} used LLMs to measure coherence of persona-driven interactions in healthcare contexts.

\textit{Internal Validity Assessment:}
A quarter of articles (25.0\%) used some form of internal validity assessment, applying LLMs to evaluate the consistency and internal structure of persona characteristics. For example, researchers used LLMs to evaluate persona-driven responses using the Big Five personality factors to understand chatbot responses to personality tests \cite{de_winter_use_2024}. Others used LLMs to interpret advertising content based on persona attributes for privacy auditing \cite{chen_why_2024}or instructed LLMs to use standardized frameworks to evaluate persona authenticity and internal consistency \cite{ji_is_2024}. Bai et al. \cite{bai_agentic_2024} used LLMs to evaluate the emotional and cognitive characteristics of the generated personas.

\textit{Statistical Validity Assessment:}
Statistical testing was applied infrequently in conjunction with LLM-based persona evaluation (16.7\% of LLM evaluations). 
Choi et al. used t-tests to compare persona quality scores between different methods \cite{choi_proxona_2025}. Cross-model studies applied statistical tests to compare lexical diversity across multiple LLMs, including GPT-4, Gemini, Claude, and Llama \cite{sethi_when_2025}. Correlation tests checked whether different LLMs produced consistent persona attributes \cite{li_steerability_2024}, whereas regression analyses can be used to test how persona characteristics affected content interpretation accuracy \cite{chen_why_2024}. These statistical approaches remain less common than qualitative validity assessments, but they do provide quantitative rigor to persona evaluation.

Pertinent to all four validity types, our analysis reveals cases in which the same LLMs are used to both generate and evaluate personas. 
This usage of LLM raises questions about evaluation independence when the same technology both produces and judges outputs, though some articles mitigate this by using different LLMs for generation versus evaluation, with the idea that another LLM is more ``independent'' in its assessment of the personas. 

\begin{table}[htbp]
\centering
\small
\caption{LLM use in persona evaluation based on validity concepts. LLMs are mostly used to assess the face credibility of personas, but are also used to assess their usage in the real world, their internal consistency, and their validity against statistics. Articles can appear in more than one category.}
\label{tab:g10-thematic-coding}
\begin{tabular}{p{3cm}p{2.5cm}p{1.5cm}}
\toprule
\textbf{Assessment Type} & \textbf{Focus Area} & \textbf{Example articles} \\
\midrule
\multirow{2}{4cm}{Face Validity \break (n = 4, 36.4\%)} & Surface Credibility & \cite{dong_can_2024}, \cite{li_iqa-eval_2024} \\
\cmidrule(lr){2-3}
 & Apparent Realism & \cite{panda_llms_2024}, \cite{rudolph_ai-based_2024} \\
\midrule
\multirow{2}{4cm}{Ecological Validity \break (n = 4, 36.4\%)} & Real-World Context & \cite{gonzalez_exploring_2024}, \cite{li_chatgpt_2024} \\
\cmidrule(lr){2-3}
 & Naturalistic Behavior & \cite{sun_kiss_2024}, \cite{xu_character_2024} \\
\midrule
\multirow{2}{4cm}{Internal Validity \break (n = 3, 27.3\%)} & Internal Consistency & \cite{ji_is_2024}, \cite{yin2025co} \\
\cmidrule(lr){2-3}
 & Structural Integrity & \cite{gonzalez_exploring_2024} \\
\midrule
\multirow{2}{4cm}{Statistical Validity \break (n = 3, 27.3\%)} & Cross-Model Validation & \cite{rudolph_ai-based_2024} \\
\cmidrule(lr){2-3}
 & Systematic Analysis & \cite{chen_why_2024}, \cite{sun_kiss_2024} \\ 
\bottomrule
\end{tabular}
\end{table}

\subsection{RQ3: What Ethical Considerations Are Associated with GenAI Personas?}

\subsubsection{Ethical Engagement Patterns}

We find that more than half (56.8\%, n = 46) of the articles explicitly discuss ethical considerations, while the rest (43.2\%, n = 35) do not. 
This pattern indicates some awareness of ethical implications in GenAI personas, although many articles still lack explicit ethical reflection.
In addition, the depth of the articles' engagement with ethical issues varies considerably from one article to another. Some articles provide comprehensive ethical frameworks that systematically address multiple concerns. These comprehensive approaches examine bias detection mechanisms, community validation requirements, and potential societal harms from misrepresenting people \cite{sun_persona-l_2025, venkit_tale_2025, doudkin_synthetic_2025}. These articles articulate specific methodologies to identify and mitigate ethical risks, such as involving community representatives in validation processes for vulnerable populations and implementing systematic bias detection protocols to identify harmful stereotypes prior to deployment.

Other articles acknowledge ethical considerations but provide a limited treatise of these issues. These articles typically mention concerns about bias, representation accuracy, or potential limitations without developing systematic frameworks for addressing them \cite{gupta_evaluation_2024, mercer2025applying, kaur_synthetic_2025}. Some articles focus primarily on technical performance metrics, while giving secondary attention to the broader implications of GenAI-based user representation.


\subsubsection{Primary Ethical Concerns} \label{sec:ethical_concerns}

In the articles, the discussions on the ethical implications of GenAI personas are characterized by four themes (see Table~\ref{tab:ethical_concerns}): (1) bias and representational concerns, (2) potential societal harms from decision making, (3) trust and validity concerns, and (4) reduced human oversight.

\textit{Bias and Representational Concerns: }
Bias and other representational concerns are a long-lasting challenge in persona development and use \cite{turner2011stereotyping,marsden_stereotypes_2016}. Our analysis reveals that researchers also express concerns about biases in GenAI personas (n = 17, 37.0\%). Systematic ``algorithmic othering'' particularly affects marginalized communities, where AI systems exhibit biases in representing gender and profession combinations \cite{venkit_tale_2025}. GenAI personas are susceptible to caricature, tending to exaggerate traits of marginalized personas; for example, depicting individuals with physical disabilities primarily as being ``in wheelchairs'' \cite{nguyen_simulating_2024}.

Gender and age biases appear in multiple contexts, including autism representation, in which personas reinforce harmful stereotypes about gender differences \cite{park_as_2025}, and financial representation in which personas do not accurately represent behaviors across age demographics \cite{kaur_synthetic_2025}. For example, Gupta et al. \cite{gupta_bias_2023} studied 24 reasoning datasets with 19 diverse personas using ChatGPT-3.5; they found 80\% of personas showed stereotypical bias. Reliance on stereotypical associations limits the generation of diverse personality combinations, creating personas of predictable character types that may not reflect real user population complexity \cite{mercer2025applying}. Platform-specific sampling biases also emerge when personas generated from specific platforms inherit the demographic biases of those data sources \cite{yin2025co}. Although these persona concerns are not new (see, e.g., \cite{turner2011stereotyping}), the introduction of AI into the persona process may exacerbate existing issues.

\textit{Potential Societal Harms from Decision Making: }
Some articles (n = 10, 21.7\%) point out that GenAI personas could cause societal harm in several ways beyond individual misrepresentation. We list here some of these ways. First, ``synthetic persuasion paradoxes'' occur when AI personas systematically overestimate intervention effects compared to real human responses, potentially leading to misguided policy decisions that assume unrealistic behavior change outcomes \cite{doudkin_synthetic_2025}. Second, discrimination perpetuation occurs when personas embed and amplify existing societal biases, particularly affecting autism and other marginalized identities \cite{park_as_2025}. This occurs when personas are used in design processes, potentially leading to products and services that inadequately serve or actively exclude affected populations.
Third, researchers note that harmful stereotyping risks emerge when generating personas for sensitive populations \cite{sun_persona-l_2025}. For example, without appropriate community validation and expert oversight, personas risk misrepresenting vulnerable groups in ways that could harm advocacy efforts and policy development. Policy misapplication represents another concern among researchers, in that biased persona data could provide overly optimistic estimates of intervention effectiveness, potentially leading to ineffective resource allocation \cite{doudkin_synthetic_2025}.

\textit{Trust and Validity Concerns:}
Other articles (n = 9, 19.6\%) express concern about the authenticity and reliability of the AI persona. We list the main concerns here. First, LLMs may yield reduced semantic diversity and more formulaic narratives compared to human-created personas \cite{venkit_tale_2025}, which raises questions about whether GenAI personas can capture the richness and complexity of actual human experiences and motivations. Second, preference misrepresentation \cite{li_consumer_2025} and issues with trust creation in user representation \cite{kaur_synthetic_2025} are other concerns in this space. As a valuable consideration, Muller and Seaborn \cite{muller_stepford_2025} refer to the use of synthetic data in persona generation leading to ``Potemkin personas''; i.e., personas that are ``staged'' to give an appearance of reality merely to ``please the eye'' rather than truthfully representing reality. For example, in addiction-focused persona research \cite{salminen_deus_2024}, GenAI consistently generated personas with 86\% US-based representation despite no geographical constraints in prompts, and portrayed addiction narratives in an unrealistically positive light, mitigating the severity of substance abuse issues.
Third, GenAI personas often lack emotional depth, which is a validity concern because emotions are an integral part of real user behavior and influence decision-making with personas \cite{li_consumer_2025}. Overall, validation gaps emerge from the absence of experimental validation with actual participants, limiting researchers' ability to assess whether GenAI personas accurately represent target populations \cite{park_as_2025}.

\textit{Reduced Human Oversight}
Our analysis shows that a major portion (42.0\%, n = 34) of the articles reported no human-AI collaboration in persona generation, while 34.6\% (n = 28) used such collaboration in a single stage (data collection, segmentation, enrichment, and evaluation) and 23.5\% (n = 19) employed collaboration at multiple stages. Human involvement occurred primarily in content writing (28.4\%) and generation (25.9\%), with minimal participation in the evaluation (16.0\%) or analysis phases.
The typical workflow for GenAI persona development involves LLMs autonomously generating personas with no or limited human oversight. When humans are involved, they mainly provide feedback and input on prompt responses \cite{paoli_improved_2023}, then review and refine LLM outputs through feedback loops \cite{zhou_vivid-persona_2024}. However, evaluation phases show reduced human involvement, despite evidence that community validation with representatives improves the representation of vulnerable populations \cite{sun_persona-l_2025} and cross-cultural validation prevents inappropriate persona transfer between contexts \cite{kaur_synthetic_2025}. 

\begin{figure}
    \centering
    \includegraphics[width=\linewidth]{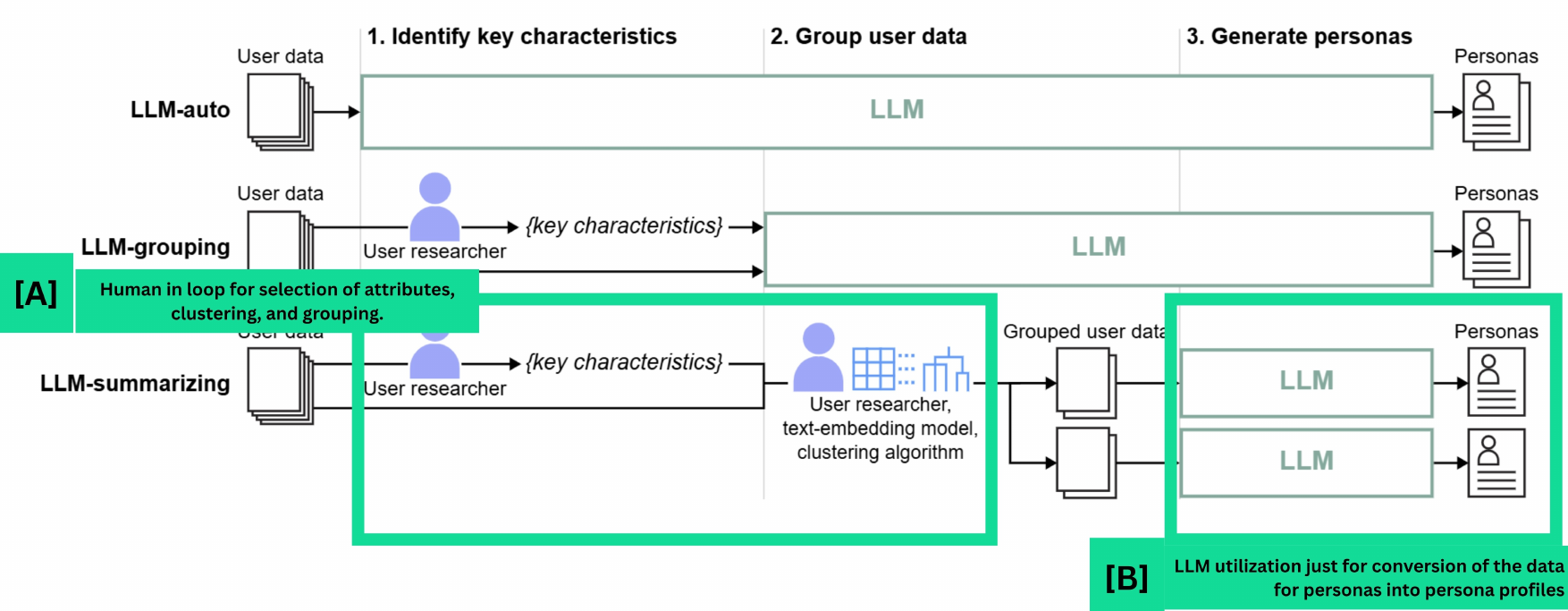}
    \caption{Collaborative workflows between humans and GenAI for persona generation \cite{shin_understanding_2024}: LLM-auto (fully automated), LLM-grouping (human researcher selects key characteristics), and LLM-summarizing with annotated sections indicating human involvement in attribute selection and clustering ([A]) and LLM utilization for data conversion ([B]). The LLM-summarizing model illustrates how AI and humans can ``collaborate'' on persona creation with humans identifying attributes and clustering while LLMs summarize data into persona narratives.The figure is annotated to explain different models of human-AI collaboration.}
    \label{fig:collaboration}
    \Description{Collaborative workflows between humans and GenAI for persona generation, showing three approaches. LLM-auto represents fully automated generation, LLM-grouping shows human researcher selection of key characteristics, and LLM-summarizing illustrates human involvement in attribute selection and clustering ([A] annotation) with LLM utilization for data conversion ([B] annotations).}
\end{figure}

\begin{table*}[htbp]
\centering
\small
\caption{Thematic analysis of ethical considerations in GenAI personas. The most common ethical concerns considered and discussed in GenAI persona literature include bias and representation, while concerns related to a lack of trust, potential harms in decision-making, and reduced human oversight are also present. See Table A9 in the online appendix for complete list of articles.}
\label{tab:ethical_concerns}
\begin{tabular}{p{4cm}p{6cm}p{3.5cm}}
\toprule
\textbf{Main Ethical Theme} & \textbf{Specific Ethical Concerns} & \textbf{Example Articles} \\
\midrule

\multirow{6}{4cm}{\textbf{Bias and Representational Concerns (n = 17, 37.0\%)}} & Algorithmic Othering & \cite{venkit_tale_2025} \\
\cmidrule(lr){2-3}
& Elite Institution Overrepresentation & \cite{gupta_evaluation_2024} \\
\cmidrule(lr){2-3}
& Gender and Age Bias & \cite{park_as_2025,kaur_synthetic_2025} \\
\cmidrule(lr){2-3}
& Stereotypical Associations & \cite{mercer2025applying} \\
\cmidrule(lr){2-3}
& Platform-Specific Bias & \cite{yin2025co} \\
\cmidrule(lr){2-3}
& AI Exacerbation of Existing Issues & \cite{park_as_2025,venkit_tale_2025} \\
\midrule

\multirow{4}{4cm}{\textbf{Potential Societal Harms from Decision Making (n = 10, 21.7\%)}} & Synthetic Persuasion Paradox & \cite{doudkin_synthetic_2025} \\
\cmidrule(lr){2-3}
& Discrimination Perpetuation & \cite{park_as_2025} \\
\cmidrule(lr){2-3}
& Harmful Stereotyping of Vulnerable Groups & \cite{sun_persona-l_2025} \\
\cmidrule(lr){2-3}
& Policy Misapplication & \cite{doudkin_synthetic_2025} \\
\midrule

\multirow{4}{4cm}{\textbf{Trust and Validity Concerns (n = 9, 19.6\%)}} & Reduced Semantic Diversity & \cite{venkit_tale_2025} \\
\cmidrule(lr){2-3}
& Consumer Preference Misrepresentation & \cite{li_consumer_2025,kaur_synthetic_2025} \\
\cmidrule(lr){2-3}
& Limited Emotional Range & \cite{li_consumer_2025} \\
\cmidrule(lr){2-3}
& Validation Gaps & \cite{park_as_2025} \\
\midrule

\multirow{5}{4cm}{\textbf{Reduced Human Oversight (n = 10, 21.7\%)}} & Limited Human-AI Collaboration & \cite{paoli_improved_2023,zhou_vivid-persona_2024} \\
\cmidrule(lr){2-3}
& Automated Generation Workflow & \cite{paoli_improved_2023,zhou_vivid-persona_2024} \\
\cmidrule(lr){2-3}
& Community Validation Requirements & \cite{sun_persona-l_2025,venkit_tale_2025} \\
\cmidrule(lr){2-3}
& Cross-Cultural Validation Gaps & \cite{kaur_synthetic_2025} \\
\cmidrule(lr){2-3}
& Evaluation Phase Neglect & \cite{paoli_improved_2023,zhou_vivid-persona_2024} \\
\bottomrule
\end{tabular}
\end{table*}

\subsubsection{Transparency and Reproducibility} \label{sec:transparency}
The field demonstrates notable advancement relative to traditional persona development practices in research openness and reproducibility. Most (61.7\%, n = 50) of the articles show commitment to open science practices by sharing at least one type of resource (personas, data, prompts, and/or code). This represents progress compared to traditional persona research, where resource sharing was exceptional \cite{salminen_literature_2020}. Among these, 16.0\% (n = 13) demonstrate comprehensive resource sharing by making all three resource types available, while 45.7\% (n = 37) share one or two resource types. Slightly more than a third (38.3\%, n = 31) of the articles share no resources at all.

Resource sharing shows encouraging patterns, with researchers sharing different types of resources. Personas represent the most commonly shared resource, with more than half (51.9\%, n = 42) of the articles making their personas publicly available. This represents a substantial improvement over previous generations of persona research, where resource sharing was exceptional \cite{salminen_literature_2020}. Code sharing shows a meaningful adoption, with nearly a third (28.4\%, n = 23) of the articles providing implementation details. 
Dataset sharing shows a similar trend, with 27.2\% of articles (n = 22) making their data accessible. 
Among the articles that share at least one resource, personas-only sharing represents the most common pattern (n = 20, 40\%), followed by combined persona and code sharing (n = 10, 20\%). Some articles demonstrate comprehensive resource sharing practices. Some articles provide persona descriptions and detailed prompts, enabling replication of their methodologies \cite{salminen_deus_2024, venkit_tale_2025}, whereas others share code repositories to facilitate the adoption and extension of GenAI use in persona development \cite{park_character-centric_2024, shin_understanding_2024}.

As a new type of reproducibility resource, prompt documentation is a particularly important aspect of GenAI persona research, which also represents a novel research area for the HCI community. According to our findings, researchers apply four main prompting strategies (see Table~\ref{tab:prompting-strategies}): (1) attribute specific instructions, (2) role-playing instructions, (3) data driven prompting, (4) creative persona development. Out of these, \textit{attribute-specific prompting} provides the highest degree of control by specifying exact demographic or behavioral characteristics through structured formats~\cite{salminen_deus_2024,kaur_synthetic_2025}, and or controlled variation of specific attributes~\cite{nguyen_simulating_2024}. In turn, \textit{role-playing instructions} take a more contextual approach, instructing the LLM to embody particular user types with specific emotional states, communication styles~\cite{barambones_chatgpt_2024,sun_persona-l_2025}, or personality traits~\cite{zhou_vivid-persona_2024,li_actions_2025}. Furthermore, \textit{data-driven prompting} grounds persona generation in actual user data, either by directly integrating survey responses, analytics, or behavioral logs into the prompts~\cite{shin_understanding_2024,jung2025personacraft}, or by using LLMs to assess which synthetic responses best match real user profiles~\cite{dong_can_2024,argyle_out_2023}. Finally, \textit{creative persona development} emphasizes psychological depth and narrative richness~\cite{park_character-centric_2024,makridis_virtualxai_2025}, often employing multimodal approaches that combine text with images or videos for persona representation~\cite{park_character-centric_2024,doudkin_synthetic_2025}. 

It is worth noting that these strategies are not mutually exclusive, but researchers often combine multiple approaches, such as using data-driven prompts to establish demographic baselines while employing role-playing instructions to add behavioral details \cite{cheng_marked_2023}. The choice of strategy depends on the research goals; for example, attribute-specific prompting applies to fully automated pipelines for persona generation, whereas role-playing is typically used to elicit richer behavioral dynamics, data-driven approaches are favored when researchers aim to anchor personas in empirical user profiles, and narrative-oriented personas are often applied in creative development contexts.

\begin{table*}[htbp]
\caption{LLM prompting strategies for persona development. LLM prompting strategies for persona generation include: (1) attribute specific prompting, where specific demographic or behavioral attributes are used to change, (2) role play based prompting, where the LLM is assigned a role, (3) data driven prompting, where the data related to the users is provided to generate personas, and (4) creative prompting which involves different modalities.}
\label{tab:prompting-strategies}

\resizebox{\textwidth}{!}{%
{\footnotesize
\begin{tabular}{p{3cm} p{3cm} p{10cm}}
\toprule
\cell{\textbf{Primary prompting strategy}} & \cell{\textbf{Specific technique}} & \cell{\textbf{Example implementation}} \\
\midrule

\multirow{6}{=}{\cell{\textbf{Attribute-specific prompting}}} &
\multirow{3}{=}{\cell{Structured format instructions}} &
\cell{``Provide the output in a JSON array, with each dict containing only the following keys: `index', `name', `age', `occupation', `background', `details''' \cite{salminen_deus_2024}} \\
\cmidrule(lr){3-3}
& & \cell{``Uses decision tree methodology to create personas with 2-8 attributes based on survey characteristics'' \cite{kaur_synthetic_2025}} \\
\cmidrule(lr){2-3}

& \multirow{3}{=}{\cell{Controlled attribute variation}} &
\cell{``Given a fictional user's description: \{base persona\} Modify the description to make it more like a \{privacy attribute\} version of the person'' \cite{nguyen_simulating_2024}} \\
\cmidrule(lr){3-3}
& & \cell{``Generate persona variants by varying selected attributes such as age and income while maintaining core demographic consistency'' \cite{nguyen_simulating_2024}} \\
\midrule

\multirow{6}{=}{\cell{\textbf{Role-playing instruction}}} &
\multirow{3}{=}{\cell{Context-specific simulation}} &
\cell{``You are a user of a university intranet... act as a young person who uses colloquial but polite language, making up a name, a gender, and an age within the range of 21 and 26 years'' \cite{barambones_chatgpt_2024}} \\
\cmidrule(lr){3-3}
& & \cell{``Uses LLM (GPT-4o mini) with RAG framework and ability-based approach for personas of people with complex needs'' \cite{sun_persona-l_2025}} \\
\cmidrule(lr){2-3}

& \multirow{3}{=}{\cell{Emotional and behavioral specification}} &
\cell{``In a conversation, you should act as if you have the following eight emotional parameters. Each parameter should fluctuate throughout the dialog...'' \cite{zhou_vivid-persona_2024}} \\
\cmidrule(lr){3-3}
& & \cell{``Uses Big Five personality traits and domain-specific profiles to create diverse novice instructor personas'' \cite{li_actions_2025}} \\
\midrule

\multirow{6}{=}{\cell{\textbf{Data-driven prompting}}} &
\multirow{3}{=}{\cell{User data integration}} &
\cell{``Here is the user data... Generate a minimum number of personas to represent the user data. Rule 1: Do not add any information that does not exist in the user data...'' \cite{shin_understanding_2024}} \\
\cmidrule(lr){3-3}
& & \cell{``PersonaCraft uses LLMs integrated with clustering analysis in a 5-stage methodology including data preparation, question mapping, pre-processing, persona generation, and enrichment phases'' \cite{jung2025personacraft}} \\
\cmidrule(lr){2-3}

& \multirow{3}{=}{\cell{Comparative assessment}} &
\cell{``Given the user profile provided below, select the response from AI assistant A or B that the user would most likely prefer...'' \cite{dong_can_2024}} \\
\cmidrule(lr){3-3}
& & \cell{``LLMs generate synthetic survey responses by conditioning on demographic personas to create digital twins of real participants'' \cite{argyle_out_2023}} \\
\midrule

\multirow{6}{=}{\cell{\textbf{Creative persona development}}} &
\multirow{3}{=}{\cell{Psychological Depth Elicitation}} &
\cell{``What is your dark secret?... What is your personality like? Personality is one of the following five traits: extraversion, agreeableness, conscientiousness, neuroticism, or openness to experience...'' \cite{park_character-centric_2024}} \\
\cmidrule(lr){3-3}
& & \cell{``Generated 1000 backstories using GPT-4o-mini, selected 100 balanced personas with diverse demographic and professional profiles for XAI evaluation'' \cite{makridis_virtualxai_2025}} \\
\cmidrule(lr){2-3}

& \multirow{3}{=}{\cell{Multimodal persona development}} &
\cell{``DALL-E 3 prompt: A character from random genre of manga... Look carefully this image, and give me your imagination of detailed description of appearance of the character...'' \cite{park_character-centric_2024}} \\
\cmidrule(lr){3-3}
& & \cell{``Automated pipeline creating synthetic profiles with demographic and behavioral attributes including names, education, careers, locations, hobbies, and climate stances'' \cite{doudkin_synthetic_2025}} \\
\bottomrule
\end{tabular}
}%
}
\end{table*}

\section{Discussion} \label{discussion}

\subsection{Research Implications} \label{research implications}

\begin{figure}
    \centering
    \includegraphics[width=\linewidth]{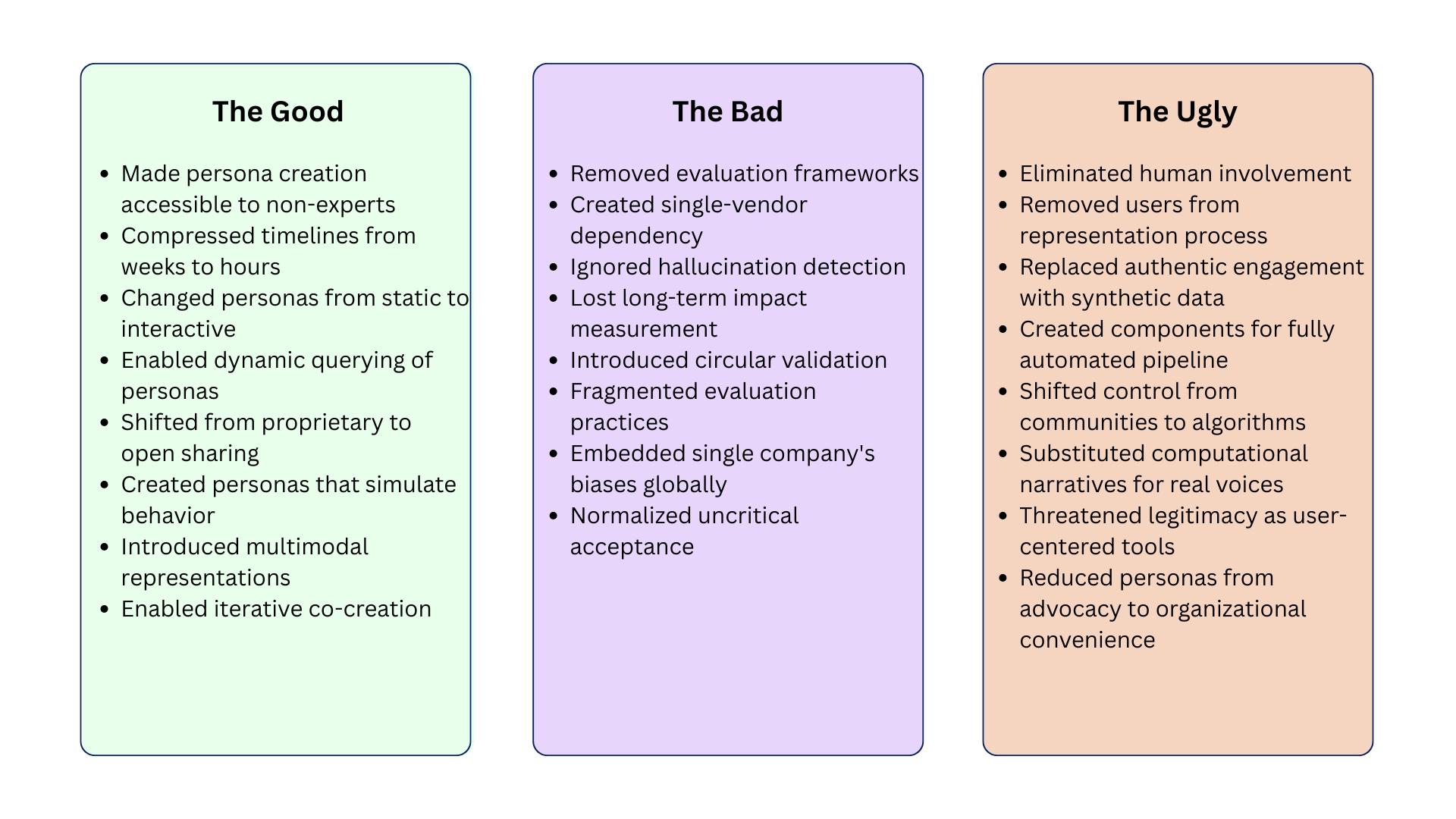}
    \caption{The good, the bad, and the ugly dimensions of the GenAI Personas.}
    \label{fig:good_bad_ugly}
    \Description{A three-panel figure showing the good, bad, and ugly trends from the research community. The good findings (colored green) include improving persona accessibility, converting static personas to interactive ones, and enhancing the open-source capabilities of the personas. The bad findings (highlighted in purple) include removing evaluation frameworks from the GenAI personas, creating a single vendor dependency, and the circular risk of evaluation by the same models used for generation. The ugly (colored orange) includes the elimination of human involvement, the use of synthetic data, and the shift of control from humans to algorithms.}
\end{figure}

To help the reader make sense of our findings better, we apply three categories:  
the Good, the Bad, and the Ugly (see Figure \ref{fig:good_bad_ugly} for elaboration). \textit{Good} refers to beneficial advances that validate the value of GenAI personas and point toward productive applications. \textit{Bad} refers to methodological weaknesses that undermine validity and reliability. \textit{Ugly} refers to systemic or structural risks that can cause harm, particularly ethical violations stemming from over-reliance on automation without human oversight.

\textit{The Good---Expanding Accessibility, Expressiveness, and Opportunities:}
It can be argued that GenAI has democratized persona development in numerous ways, such as increasing persona generation speed and enabling researchers without specialized UX training to create personas. GenAI persona research is also more likely to share resources than previous eras of persona research, which aligns with broader movements towards open science practices and ideals like increased transparency and reproducibility. Moreover, GenAI enables new interaction formats, such as conversational personas that allow designers to interact with personas dynamically rather than read static profiles. There are also new, \textit{iterative co-creation} models where humans and machines collaborate through continuous prompt refinement. Moreover, GenAI enables \textit{agentic personas} that simulate user responses, which challenges the traditional distinction of personas primarily describing user segments by broadening the scope of personas into predicting user behaviors. While traditional personas serve as reference documents that designers consulted to understand user needs \cite{pruitt2010persona,cooper1999inmates}, agentic personas generate responses representing user views in real time. This raises an intriguing question, which is: \textit{Is a persona legitimate because it accurately describes real users based on data, or because it can accurately predict what users would do?} This is an exciting and potentially impactful area for HCI research. 

\textit{The Bad---The Accessibility-Quality Paradox:}
However, there is also what we term the \textit{accessibility-quality paradox: even though GenAI can democratize persona creation, its precarious use can severely undermine methodological rigor}. 
GenAI technology lowers the skill barriers that prevent untrained creators from creating personas, but evaluation still requires expertise in validity assessment and bias detection \cite{salminen_persona_2020,chapman_personas_2006}. When anyone can generate personas in minutes, but proper evaluation requires significant expertise, persona developers are incentivized to treat evaluation light-heartedly. This can explain why generation techniques exhibit a degree of standardization, but evaluation remains more inconsistent among studies. The problem mirrors longstanding concerns that personas lack empirical validation \cite{Jansen2021}, but GenAI can be seen to erode some of the quality gatekeeping that technical barriers previously provided.
In GenAI persona research, we observe hints of this pattern. Nearly half (44\%) of articles lack explicit evaluation frameworks, weakening the empirical foundation of persona research. The evaluation challenge is exacerbated by the fact that only 11.5\% of the articles explicitly address hallucination detection, even though this risk is common knowledge \cite{hamalainen_evaluating_2023}. 
In general, the evaluation of GenAI personas varies considerably in its degree of human involvement. Human-in-the-loop approaches employ people to assess personas at specific stages \cite{jung2025personacraft}. In contrast, automated evaluation pipelines integrate computational assessments directly into the generation workflow, applying NLP metrics, stereotype detection algorithms, or LLM-based judges without human intervention, thereby reducing transparency \cite{cheng_marked_2023,sethi_when_2025}.

A particular risk occurs when the same LLM generates and evaluates personas, risking that the evaluation is based on model-specific patterns rather than the genuine quality of the personas. Unlike persona generation, which has converged on common patterns such as prompt engineering and data grounding, evaluation practices for GenAI personas are currently \textit{ad hoc}. The field lacks consensus on when evaluation should occur, who should conduct the evaluation, and what constitutes adequate validation. Also, most research (86.4\%) depends on a single provider (OpenAI's GPT models), which limits diversity and embeds one company's biases in persona development, such as embedding Western perspectives into global persona representations \cite{atari_which_2023}. This concentration exemplifies over-reliance on automated systems at the expense of critical evaluation \cite{skitka2000automation}. The reliance on a single provider may lead to uncritical acceptance of GenAI personas, particularly when practitioners lack the technical expertise to evaluate LLM limitations. These problems could be addressed through standardized evaluation protocols, diversified LLM use, and mechanisms to detect hallucinations.

\textit{The Ugly---Loss of Human Agency and Participatory Principles:}
Some developments threaten the foundation of persona research itself by systematically reducing human involvement. The patterns we observe in GenAI persona research are problematic from the perspective of participatory design. Nearly half (42.0\%) of the articles do not report an explicit human-AI collaboration model, instead fully delegating the persona development process to GenAI. About a third (30.9\%) of the articles adopt participatory methods. This creates a risk of computational efficiency substituting genuine engagement, producing personas that lack authentic user voices. Our research shows evidence of each step toward fully automated pipelines. We found articles that used synthetic data for persona generation, articles that used LLMs only to generate personas, but others that also evaluated personas using LLM, and some articles that presented complete persona systems that produce autonomous decisions without human intervention.

In the worst scenario, one could perceive a dystopian outcome in which personas are created from synthetic data by machines that are then evaluated by other machines, and possibly delivered to yet other machines making decisions about real people. This scenario would eradicate humans from the design process and goes against the human-in-the-loop principles that argue for maintaining human agency and oversight in GenAI systems, which is essential when making decisions affecting people \cite{mosqueira2023human}. The fully automated persona pipeline could represent precisely what human-in-the-loop frameworks seek to prevent: algorithmic systems making consequential decisions about human populations without meaningful human involvement or accountability mechanisms. Fully automated pipelines break what makes personas valuable. Personas work in user-centered design because they represent real users' needs and perspectives \cite{pruitt2010persona,schuler1993participatory}. But the pattern we observed creates a closed loop: synthetic data feeds AI generation, AI evaluates its own outputs, and AI makes decisions about real people. No actual users are involved. This transforms personas from user representations into algorithmic artifacts with no connection to the users they claim to represent. Such systems violate basic participatory design principles that require involving those affected by design decisions \cite{schuler1993participatory,mosqueira2023human}.

Simultaneously, the increased popularization of GenAI personas reflects changing power dynamics in who (or what) controls user representation. Whereas personas initially emerged from direct user research and community engagement, GenAI personas may create computationally generated narratives that mimic authentic UX \cite{kapania2025simulacrum}. This mimicry raises epistemic questions about the trade-off between efficiency gains and representational authenticity \cite{muller_stepford_2025}, which are questions that the HCI community is required to address. These findings align with concerns in human-AI collaboration research about automation bias, where over-reliance on AI systems leads to uncritical acceptance of outputs even when human evaluation could detect errors \cite{skitka2000automation}. The patterns we observe contradict human-centered AI principles that emphasize maintaining human agency and meaningful oversight in systems that affect real populations \cite{shneiderman_human-centered_2022,mosqueira2023human}. 

Recognizing these differences can help the HCI community embrace beneficial applications of GenAI in personas while avoiding practices that would disconnect personas from the users they claim to represent. The HCI community can make a substantial contribution in this area by ensuring, whenever possible, to engage real users and maintain the participatory foundations that give personas their validity and value.

\subsection{Extension of Prior Persona Research}

Here, we position GenAI personas in the broader HCI literature on personas. We do so by first discussing the parallels between GenAI persona research and HCI research evolution, and then focusing on the continuum of data-driven personas, specifically how GenAI personas are positioned within it.

Go et al. \cite{goh2017riding} investigated the ``three waves of personas'' by reviewing 315 articles that span 18 years of research tradition. There are interesting parallels in this most recent transition of persona technology in regards to these waves. In Bødker's terms \cite{bodker2006second,bodker2015third}, there was a \textit{transition from human factor to human actor} in HCI; however, for persona creation, the adoption of LLMs seems to decrease human agency in the persona generation process. Bødker also identified \textit{engagement of other disciplines in HCI research}; this is evident in personas becoming increasingly a nexus of HCI-NLP collaboration, rather than being ``owned and operated'' by HCI alone. This idea has previously been articulated by Duda \cite{duda2018personas}, who questioned whether personas are conceptually claimed by HCI or the field of marketing; in the GenAI persona era, this question particularly relates to the persona concept being appropriated by NLP researchers. NLP research tends to have a technical tradition but less knowledge about the conceptual heritage of personas in HCI, which implies that HCI researchers' knowledge about persona theory could contribute to better operationalization of ``NLP-based'' personas (e.g., increasing persona depth and arguing for grounding on real user experiences). 

An important aspect of Bødker's \cite{bodker2006second,bodker2015third} articulation is the role of \textit{participatory design}, in which end-users are actively involved as co-designers throughout the design process. Our findings show that this is relatively rare in GenAI personas; however, it need not be rare. The democratization effect of GenAI technologies in making persona generation easier could, in theory, facilitate and support the inclusion of persona users in the generation process. Mobilizing this ``participatory potential'' remains a highly impactful avenue for future research. It is not a novel challenge; already in their 2017 review, Goh et al. \cite{goh2017riding} identified a ``a missing link between personas and designers'' (p. 612). 

Similarly, it is worth noting that these challenges, such as stereotyping, marginalization, and personas being disconnected from real user data are well-known in the persona field \cite{turner2011stereotyping,marsden_stereotypes_2016,goh2017riding} but it is nonetheless interesting to observe that they remain topical for GenAI personas. Qualitative user data and participatory validation, followed by persona refinement, are foundational methods used in persona creation. These approaches emphasize methodological transparency and stakeholder engagement~\cite{cooper1999inmates, pruitt2003personas}. As such, validation has always been central to personas, with UCD guidelines highlighting the need for empirical grounding, data triangulation, and participatory checks with both designers and users \cite{miaskiewicz2011personas, arhippainen2013tutorial, pruitt2010persona}. Our findings indicate that GenAI personas frequently lack validation, with many of the studies we examined reporting no human supervision or explicit evaluation frameworks. Thus, GenAI amplifies concerns about persona validity by accelerating production but eroding safeguards. We refer the reader to the article by Amin et al. \cite{amin2025generative} that lists twenty challenges of GenAI personas in HCI. The resolution of these challenges remains an ongoing research effort.

In their literature review spanning 15 years of research on data-driven personas, Salminen et al. \cite{salminen2021survey} argued for three evolutionary periods: (1) \textit{quantification} (2005-2008), which involved early experimentation with statistical and ML technologies in persona creation; (2) \textit{diversification} (2009-2014), which involved expanding the scope of algorithms and data sources to involve NLP and both structured and unstructured data; and (3) \textit{digitalization} (2015-2020) that marked the rise of fully automated persona generation in user-friendly systems. The review also specifically pointed out challenges in lacking shared resources, standardized evaluation methods, the risk of not generating inclusive personas, and the over-focus on algorithms at the expense of human experience. 

These risks do remain relevant for GenAI personas, but there is also progress, especially in the sharing of computational resources, using quantitative evaluation, and the researchers becoming more aware of issues such as the need for representing marginalized users and the role of algorithmic bias in persona generation. So, it appears that personas are ``moving forward'', albeit slowly. It is challenging to propose an all-encompassing label for the current era of persona generation, but perhaps (4) \textit{augmentation} could be descriptive, with the undertone that GenAI technologies exist to augment human involvement in persona generation rather than replacing it.

\subsection{Practical Implications}

Based on our analysis of 81 articles, we propose practical guidelines for the development of GenAI personas (Table~\ref{tab:practical-guidelines}). Each guideline addresses specific methodological gaps identified in our review and is grounded in relevant research from our scoping review and previous literature. Each guideline in Table~\ref{tab:practical-guidelines} addresses a specific gap from our review, labeled as \texttt{Review Finding}, and builds on prior research, labeled as \texttt{Prior Work}, with the \texttt{Reasoning} column explaining how the cited work supports the guideline. For example, PG1 corresponds to our finding that 86\% of articles rely exclusively on GPT models (Section 4.1.1), while prior work demonstrates that model-specific biases exist that single-model approaches might miss.

\begin{table*}[htbp]
\centering
\small
\caption{Practical guidelines for GenAI persona development.}
\label{tab:practical-guidelines}
\begin{tabular}{p{5.5cm}p{2.3cm}p{2cm}p{4cm}}
\toprule
\textbf{Guideline} & \textbf{Review Finding} & \textbf{Prior Work} & \textbf{Reasoning} \\
\midrule
\textbf{PG1: Implement multi-model validation using at least two different GenAI systems} & See Section \ref{sec:tech_adoption} & \cite{kosch2024risk,gupta_bias_2023,cheng_marked_2023} & Different models exhibit distinct biases that cross-validation can identify. \\
\midrule
\textbf{PG2: Deploy structured formats (JSON, XML, or CSV) with explicit consistency criteria} & See Section \ref{sec:data_sources} & \cite{jung2025personacraft,salminen_deus_2024} & Structured formats enable programmatic validation and cross-system compatibility. \\
\midrule
\textbf{PG3: Document complete methodologies including prompts, parameters, and model versions} & See Section \ref{sec:transparency} & \cite{kosch2024risk,prpa_challenges_2024} & Prompt sensitivity and parameter choices significantly affect outputs and reproducibility. \\
\midrule
\textbf{PG4: Validate generated characteristics against demographic benchmarks or real user data} & See Section \ref{sec:eval_methods} & \cite{li_llm_2025,atari_which_2023} & LLM-generated personas show systematic deviations from real-world population distributions. \\
\midrule
\textbf{PG5: Establish quantitative acceptance thresholds before incorporating personas into design} & See Section \ref{sec:eval_methods} & \cite{salminen_persona_2020,jung2025personacraft} & Pre-registered criteria prevent confirmation bias and improve persona quality. \\
\midrule
\textbf{PG6: Assess output consistency through repeated generation and edge case testing} & See Section \ref{sec:ethical_concerns} & \cite{kaate_fourth_2025,gupta_evaluation_2024,park_as_2025} & High variability indicates unstable representations and edge cases reveal model limitations. \\
\midrule
\textbf{PG7: Implement systematic human oversight for cultural appropriateness and stereotype detection} & See Section \ref{sec:ethical_concerns} & \cite{venkit_tale_2025,marsden_stereotypes_2016,turner2011stereotyping} & Automated methods miss cultural details that require human domain expertise. \\
\bottomrule
\end{tabular}
\end{table*}

These guidelines address three key ethical concerns about using GenAI personas in HCI and design. 
First, multi-model validation and consistency testing address systematic biases, inconsistencies, and harmful stereotypes that could compromise research validity and perpetuate discrimination. These validation approaches identify systematic issues while maintaining stable characteristics across iterations, addressing concerns about reproducibility in persona development. Second, empirical grounding through structured outputs, demographic validation, and quantitative thresholds ensures that GenAI personas remain anchored in real user data rather than generating fictional or biased representations based solely on training patterns or artifacts. For example, requiring personas to match validated demographic distributions prevents models from creating representations that diverge from actual user populations. 

Third,  documentation requirements and human oversight protocols provide ethical safeguards against cultural insensitivity or stereotyping, because GenAI personas may lack an intricate understanding of cultural contexts that human researchers possess. Overall, these guidelines aim at leveraging GenAI's efficiency in persona generation while maintaining the methodological rigor and ethical standards expected in HCI research, so that GenAI personas offer valid, reliable, and respectful representations of diverse user populations. Moreover, we expect that human-AI collaboration models will become increasingly important with the rise of agentic personas that are capable of generating their own outputs independently.


\subsection{Limitations and Future Research}

Our study has some limitations. The on-going formation of practices in the GenAI persona research means that our analysis represents a snapshot of a rapidly changing field. The dominance of recent publications (88.9\% from 2024-2025) reflects the field's nascency, but limits assessment of long-term trends. However, scoping reviews are valuable in such cases, and we do share our resources (search query, coding book, corpus) so that it can be updated in future. 

One limitation is that our HCI-focused perspective may not capture relevant developments in adjacent fields such as NLP or computational social science, as we excluded articles that use ``persona'' to configure LLM personalities or agent behaviors \cite{tao2024rolecraft,li_steerability_2024,kosenko2024krgp,kamruzzaman2024woman}. Although outside of our scope, these agent-focused persona studies may offer transferable insights into prompt engineering techniques, personality consistency methods, or behavioral alignment approaches. Future research could examine the synergies between the HCI and NLP persona research traditions.

In addition to our research limitations, the reviewed articles indicate recurring challenges that constrain the field's development and can guide future work. For example, token processing constraints appear frequently in complex persona generation \cite{paoli_improved_2023}, limiting the depth and complexity of GenAI personas due to technical limitations in current LLM architectures. \textit{Bias and representation} issues are prevalent in researchers' treatise of GenAI personas. For example, Kaur et al. \cite{kaur_synthetic_2025} documented systematic biases that affect cross-cultural applicability, while Mercer et al. \cite{mercer2025applying} identified a heavy reliance on stereotypical associations that constrain persona diversity. Yin et al. \cite{yin2025co} found platform dependency issues that introduce systematic sampling biases based on data source characteristics. \textit{Validation challenges} include small sample sizes that limit generalizability \cite{venkit_tale_2025}, and the absence of experimental validation with actual participants \cite{park_as_2025}. 
\textit{Cultural limitations} represent another significant constraint. Studies confined to specific regions limit cross-cultural generalizability, raising questions about whether GenAI persona methods in one cultural context can be appropriately transferred to others.

Future research directions should address the limitations identified above, offering significant research fronts for the HCI community. First work on \textit{bias mitigation} is needed because current articles reveal persistent issues related to representational bias, stereotyping, and platform-dependent sampling. For example, systematic bias–reduction strategies \cite{doudkin_synthetic_2025} and community-based validation practices \cite{venkit_tale_2025} offer promising starting points for improving cross-cultural reliability. Second,  technical improvements are required to overcome LLM design constraints that limit persona depth and generalizability. For example, multi-platform integration \cite{yin2025co} can reduce single-platform dependence, multimodal pipelines \cite{makridis_virtualxai_2025} can support more accessible and usable persona representations, and domain-specific fine-tuning \cite{li_consumer_2025} could help improve the validity of personas in specialized contexts. 
Third, the validity and long-term utility of GenAI personas should be studied. For example, longitudinal study designs \cite{sun_persona-l_2025} can assess the stability of personas over time, stereotype-detection mechanisms \cite{gonzalez_exploring_2024} can help monitor problematic persona narratives, and cross-cultural validation frameworks \cite{kaur_synthetic_2025} are needed to determine whether GenAI technologies and techniques developed in one region transfer to others. 

\section{Conclusion}

GenAI technology is increasingly adopted for persona generation, but the field remains fragmented and methodologically inconsistent. The articles we examined demonstrate progress in prompt engineering, multimodal generation, resource sharing, and domain-specific adaptation, but they also reveal recurring limitations related to bias, validation, and cultural transferability. Our findings suggest that the main challenge for HCI is not producing more complex and automated GenAI persona pipelines, but validating that GenAI personas are reliable, interpretable, involve humans in development, and appropriate for real use. Overall, the field can benefit from addressing the specific risks and limitations identified in this review. Through such efforts, GenAI personas can mature from a technological novelty into a robust, and socially beneficial research approach.

\bibliographystyle{ACM-Reference-Format}
\bibliography{references}
\end{document}